\documentclass[11pt,a4paper,reqno]{amsart}
\usepackage[hmargin={2.7cm,2.7cm},vmargin={2.5cm,2.5cm},centering]{geometry}
\usepackage{amssymb,amsmath,amsthm,amsfonts,amscd}
\usepackage{graphicx}
\usepackage[dvipsnames]{xcolor}
\usepackage{mathrsfs}
\usepackage[unicode,psdextra]{hyperref}
\usepackage[utf8]{inputenc}
\hypersetup{pdfborder={0 0
    0.06},citebordercolor=[rgb]{0.8196,0.2275,0.5098},linkbordercolor=[rgb]{0.1765,0.5490,0.8118},urlbordercolor=[rgb]{0.7059,0.5333,0.1137}}
\usepackage{tikz}
\usetikzlibrary{decorations.markings}
\usepackage{tikz-cd}
\usepackage[english]{babel}
\usepackage{dsfont}
\usepackage{url}
\usepackage[longnamesfirst,numbers,sort&compress]{natbib}

\newcount\driver
\newcount\bozza

\font\ottorm=cmr8 scaled\magstep1 
scaled\magstep1                  
scaled\magstep1 \font\msytw=msbm10 scaled\magstep1
 
scaled\magstep1                 \font\indbf=cmbx10 scaled\magstep2

scaled \magstep2

{\count255=\time\divide\count255 by 60
\xdef\hourmin{\number\count255}
        \multiply\count255 by-60\advance\count255 by\time
   \xdef\hourmin{\hourmin:\ifnum\count255<10 0\fi\the\count255}}

\let\a=\alpha \let\b=\beta    \let\g=\gamma     \let\d=\delta     \let\e=\varepsilon
  \let\h=\eta     \let\th=\vartheta      \let\l=\lambda
\let\m=\mu    \let\n=\nu      \let\x=\xi                \let\r=\rho
\let\s=\sigma \let\t=\tau            
\let\ps=\psi        
\let\G=\Gamma        \let\L=\Lambda

\def\PP{{\mathcal P}}\def\EE{{\mathcal E}}\def\VV{{\mathcal V}}
\def\WW{{\mathcal W}}
\def\TT{{\mathcal T}}\def\NN{{\mathcal N}}\def\BB{{\mathcal B}}\def\ZZ{{\mathcal Z}}
\def\RR{{\mathcal R}}\def\LL{{\mathcal L}}

\def\xx{{\bf x}}
  
\def\kk{{\bf k}}\def\nn{{\bf n}}

 \def\bP{{\bf P}}

       \def\oo{{\underline \omega}}
\def\ee{{\underline \varepsilon}}

\def\RRR{\hbox{\msytw R}}

\def\NNN{\hbox{\msytw N}}          
        \def\ZZZ{\hbox{\msytw Z}}

        \def\EE{\hbox{\msytw E}}

\let\bs=\backslash
\let\==\equiv

\let\io=\infty
\let\0=\noindent

\def\*{{\hfill\break\null\hfill\break}}

\def\tilde#1{{\widetilde #1}}

\def\tende#1{\,\vtop{\ialign{##\crcr\rightarrowfill\crcr
             \noalign{\kern-1pt\nointerlineskip}
             \hskip3.pt${\scriptstyle #1}$\hskip3.pt\crcr}}\,}
\def\otto{\,{\kern-1.truept\leftarrow\kern-5.truept\to\kern-1.truept}\,}

\def\wh#1{\widehat{#1}}
\def\hat#1{\wh{#1}}
\def\sqt[#1]#2{\root #1\of {#2}}

\def\bp{{\bar \ps}}

\usepackage{oldlfont}

\def\T#1{{#1_{\kern-3pt\lower7pt\hbox{$\widetilde{}$}}\kern3pt}}
\def\VVV#1{{\underline #1}_{\kern-3pt
\lower7pt\hbox{$\widetilde{}$}}\kern3pt\,}
\def\W#1{#1_{\kern-3pt\lower7.5pt\hbox{$\widetilde{}$}}\kern2pt\,}

\def\indica{\leaders \hbox to 0.5cm{\hss.\hss}\hfill}
\def\guida{\leaders\hbox to 1em{\hss.\hss}\hfill}
\mathchardef\oo= "0521

\def\xx{{\bf x}}
\def\kk{{\bf k}}\def\nn{{\bf n}}

 \def\bP{{\bf P}}
 
\def\oo{{\underline \omega}}

\def\qed{\raise1pt\hbox{\vrule height5pt width5pt depth0pt}}
 
  \def\bp{{\bar p}} 

\def\indic{\hbox{\raise-2pt \hbox{\indbf 1}}}

\def\RRR{\hbox{\msytw R}} 
 
\def\NNN{\hbox{\msytw N}} 
 \def\ZZZ{\hbox{\msytw Z}}

\def\ins#1#2#3{\vbox to0pt{\kern-#2 \hbox{\kern#1 #3}\vss}\nointerlineskip}

\newdimen\xshift \newdimen\xwidth \newdimen\yshift

\def\insertplot#1#2#3#4#5#6{%
\xwidth=#1pt \xshift=\hsize \advance\xshift by-\xwidth \divide\xshift by 2%
\begin{figure}[ht]
\vspace{#2pt} \hspace{\xshift}
\begin{minipage}{#1pt}
#3 \ifnum\driver=1 \griglia=#6
\ifnum\griglia=1 \openout13=griglia.ps \write13{gsave .2
setlinewidth} \write13{0 10 #1 {dup 0 moveto #2 lineto } for}
\write13{0 10 #2 {dup 0 exch moveto #1 exch lineto } for}
\write13{stroke} \write13{.5 setlinewidth} \write13{0 50 #1 {dup 0
moveto #2 lineto } for} \write13{0 50 #2 {dup 0 exch moveto #1
exch lineto } for} \write13{stroke grestore} \closeout13
\includegraphics{griglia.ps} \fi
\includegraphics{#4.ps}\fi%
\ifnum\driver=2 \fi
\end{minipage}
\caption{#5}
\end{figure}
}

\newdimen\shift \shift=-1.5truecm
\def\lb#1{%
\ifnum\bozza=1
\label{#1}\rlap{\hbox{\hskip\shift$\scriptstyle#1$}}
\else\label{#1} \fi}

\def\be{\begin{equation}}
\def\ee{\end{equation}}
\def\bea{\begin{eqnarray}}\def\eea{\end{eqnarray}}
\def\bean{\begin{eqnarray*}}\def\eean{\end{eqnarray*}}
\def\bfr{\begin{flushright}}\def\efr{\end{flushright}}
\def\bc{\begin{center}}\def\ec{\end{center}}
\def\bal{\begin{align}}\def\eal{\end{align}}
\def\ba#1{\begin{array}{#1}} \def\ea{\end{array}}
\def\bd{\begin{description}}\def\ed{\end{description}}
\def\bv{\begin{verbatim}}\def\ev{\end{verbatim}}
\def\nn{\nonumber}
\def\Halmos{\hfill\vrule height10pt width4pt depth2pt \par\hbox to \hsize{}}
\def\pref#1{(\ref{#1})}

\def\ins#1#2#3{\vbox to0pt{\kern-#2 \hbox{\kern#1 #3}\vss}\nointerlineskip}

\newdimen\xshift \newdimen\xwidth \newdimen\yshift
\newcount\griglia

\def\insertplot#1#2#3#4#5#6{%
\xwidth=#1pt \xshift=\hsize \advance\xshift by-\xwidth \divide\xshift by 2%
\begin{figure}[ht]
\vspace{#2pt} \hspace{\xshift}
\begin{minipage}{#1pt}
#3 \ifnum\driver=1 \griglia=#6
\ifnum\griglia=1 \openout13=griglia.ps \write13{gsave .2
setlinewidth} \write13{0 10 #1 {dup 0 moveto #2 lineto } for}
\write13{0 10 #2 {dup 0 exch moveto #1 exch lineto } for}
\write13{stroke} \write13{.5 setlinewidth} \write13{0 50 #1 {dup 0
moveto #2 lineto } for} \write13{0 50 #2 {dup 0 exch moveto #1
exch lineto } for} \write13{stroke grestore} \closeout13
\includegraphics{griglia.ps} \fi
\includegraphics{#4.ps}\fi%
\ifnum\driver=2 \fi
\end{minipage}
\caption{#5}
\end{figure}
}

\newdimen\shift \shift=-1.5truecm
\def\lb#1{%
\label{#1}\rlap{\hbox{\hskip\shift$\scriptstyle#1$}}
\else\label{#1} \fi}

\def\be{\begin{equation}}
\def\ee{\end{equation}}
\def\bea{\begin{eqnarray}}\def\eea{\end{eqnarray}}
\def\bean{\begin{eqnarray*}}\def\eean{\end{eqnarray*}}
\def\bfr{\begin{flushright}}\def\efr{\end{flushright}}
\def\bc{\begin{center}}\def\ec{\end{center}}
\def\bal{\begin{align}}\def\eal{\end{align}}
\def\ba#1{\begin{array}{#1}} \def\ea{\end{array}}
\def\bd{\begin{description}}\def\ed{\end{description}}
\def\bv{\begin{verbatim}}\def\ev{\end{verbatim}}
\def\nn{\nonumber}
\def\Halmos{\hfill\vrule height10pt width4pt depth2pt \par\hbox to \hsize{}}
\def\pref#1{(\ref{#1})}

\usepackage{amsmath}
\usepackage{amsfonts}
\usepackage{amssymb}
\usepackage{epstopdf}

\font\msytw=msbm9 scaled\magstep1 
scaled\magstep1

\let\a=\alpha \let\b=\beta  \let\g=\gamma  \let\d=\delta
\let\e=\varepsilon
  \let\h=\eta   \let\th=\theta  \let\l=\lambda
\let\m=\mu    \let\n=\nu    \let\x=\xi         \let\r=\rho
\let\s=\sigma \let\t=\tau    
\let\ps=\Psi   
\let\G=\Gamma   \let\L=\Lambda

\def\EE{{\cal E}} \def\VV{{\cal V}}
 \def\WW{{\cal W}}
\def\TT{{\cal T}}\def\NN{{\cal N}} 
\def\RR{{\cal R}}\def\LL{{\cal L}}

 \def\xx{{\bf x}}  
\def\kk{{\bf k}}
\def\PP{{\bf P}}

\def\nn{\nonumber}

\def\RRR{\hbox{\msytw R}} 

\def\NNN{\hbox{\msytw N}} 
 \def\ZZZ{\hbox{\msytw Z}}

\def\\{\hfill\break}
\def\={:=}
\let\io=\infty
\let\0=\noindent

\def\tende#1{\,\vtop{\ialign{##\crcr\rightarrowfill\crcr\noalign{\kern-1pt
    \nointerlineskip} \hskip3.pt${\scriptstyle #1}$\hskip3.pt\crcr}}\,}
\def\otto{\,{\kern-1.truept\leftarrow\kern-5.truept\to\kern-1.truept}\,}

\def\wh{\widehat}
\def\to{\rightarrow}

\def\qed{\hfill\raise1pt\hbox{\vrule height5pt width5pt depth0pt}}

\def\be{\begin{equation}}
\def\ee{\end{equation}}
\def\bp{\begin{pmatrix}}
\def\ep{\end{pmatrix}}
\def\bea{\begin{eqnarray}}
\def\eea{\end{eqnarray}}
\def\nn{\nonumber}
\def\pref#1{(\ref{#1})}

\def\lb{\label}

\newtheorem{lemma}{Lemma}[section]
\newtheorem{theorem}{Theorem}[section]

\begin{document}

\title{Vanishing of the anomaly in lattice chiral gauge theory}

\author[V.\ Mastropietro]{Vieri Mastropietro}

\address{Università degli Studi di Milano\\Dipartimento di Matematica\\Via
  Saldini 50\\20133 Milano\\Italy}

\email{vieri.mastropietro@unimi.it}


\maketitle
\begin{abstract} 
The anomaly cancellation is a basic property of the Standard Model, crucial for its consistence. 
We consider a lattice chiral gauge theory of massless Wilson fermions interacting with 
a non-compact massive $U(1) $ field coupled with left and right handed fermions in four dimensions. We prove in the infinite volume limit, for weak coupling
and inverse lattice step of the order of boson mass, that the anomaly
vanishes
up to subleading corrections and  under the same condition as in the continuum.
The proof is based on a combination 
of exact Renormalization Group, non perturbative decay bounds of correlations and lattice symmetries.
\end{abstract} 
\maketitle


\section{Introduction and Main results}

\subsection{Chiral gauge theory}

The perturbative consistence (renormalizability) of the Standard Model
relies on the vanishing of the anomalies, achieved under certain algebraic conditions
\cite{B111} severely constraining the elementary particles charges and providing a partial explanation of the charge quantization. 
In order to go beyond 
a purely perturbative framework in terms of diverging series
\cite{Se}, one needs a lattice formulation with
functional integrals
with cut-off much higher than the experiments scale;
due to triviality \cite{F}, \cite{AD}, the cut-off cannot be completely removed, at least in the Electroweak sector, hence the theory
can be seen as an effective one. 

One expects a relation between the perturbative renormalizability properties and the size of the cut-off. 
The electroweak theory is renormalizable \cite{W},\cite{TH}  so that 
a construction up to exponentially large cut-off could be in principle possible, and such cut-off is much higher than the scales of experiments.
However, this requires as a crucial prerequisite that the 
anomalies cancel, at least to a certain extent. This rises the natural question: 
does the anomaly cancel at a non-perturbative level with finite lattice, 
under the same condition as in the continuum?

In the continuum, the cancellation is based on compensations at every order
\cite{AB} based on dimensional regularizations and symmetries, but
finite lattice cut-off produce corrections and the question is if they cancel or not.
Jacobian arguments are used to support vanishing of higher orders contributions to anomalies but are essentially one loop results, as shown in 
\cite{K}.
Topological arguments explain
the anomaly cancellation on a lattice
\cite{Z} with classical gauge fields,
but in the quantum case they work only at lowest order (one loop).
The cancellation
would be obtained if a non-perturbative 
regulator for lattice chiral gauge theories could be found, but this is a long 
standing unsolved problem and only order by order results are known \cite{B4},\cite{B3}.

We consider a lattice chiral gauge theory, given by $2N$ massless fermions in four dimensions, labeled by an index $i=1,...,2N$;
we also define the indices $i_1=1,...,N$ and $i_2=N+1,...,2N$.
If the gamma matrices are
\be \g_0= \begin{pmatrix} 0 & I \\ I &0 \end{pmatrix}\quad \g_j= \begin{pmatrix} 0 & i\s_j \\-i\s_j &0 \end{pmatrix}, \quad\g_5=\begin{pmatrix}&I&0\\
          &0&-I\end{pmatrix}\ee
and $\s_\m^L=(\s_0,i \s) $, $\s_\m^R=(\s_0,-i \s) $, 
\be \s_1=\begin{pmatrix}&0&1\\
          &1&0\end{pmatrix}
 \quad
\s_2=\begin{pmatrix}
&0&-i\\ &i&0
\end{pmatrix}
\quad\s_3=\begin{pmatrix}&1&0\\
          &0&-1\end{pmatrix}\ee
the formal continuum action is given by the following expression:
\bea
&&\int dx F_{\m,\n}F_{\m,\n} +  \sum_{i_1} \int dx 
[\psi^+_{i_1, L,x} \s_\m^L (\partial_\m+\l  Q_{i_1} 
 A_\m)
\psi^-_{i_1, L,x}+\psi^+_{i_1, R,x} \s_\m^R \partial_\m
\psi^-_{i_1, L,x}]
\nn\\
&&\sum_{i_2} \int dx [
\psi^+_{i_2, R,x} \s_\m^R (\partial_\m+\l  Q_{i_2} 
 A_\m)
\psi^-_{i_2, R,x}+\psi^+_{i_2, L,x} \s_\m^L \partial_\m
\psi^-_{i_2, L,x}]
\label{h111}\eea
with $\m=(0,1,2,3)$ and $F_{\m\n}=\partial_\m A_\n-\partial_\n A_\m$.
Note that the $R$ fermions of kind $i_1$ and 
the $L$ fermions of kind $i_2$ decouple and
are
fictitious, non interacting degrees of freedom, which are convenient to introduce in view of the lattice 
regularization, see eg \cite{B3c}, \cite{B3d}.The total current coupled to $A_\m$ is 
\be
j^T_\m=\sum_{i_1} Q_{i_1}
\psi^+_{i_1, L,x} \s_\m^L
\psi^-_{i_1, L,x}+\sum_{i_2} Q_{i_2}
\psi^+_{i_2, R,x} \s_\m^R 
\psi^-_{i_2, R,x}
\ee
and the axial and vector part of the current is
\be
j^{T,V}_\m={1\over 2}\sum_{i} Q_i j_{\m,i,x}\quad\quad j^{T,A}_\m={1\over 2}\sum_{i} Q_i \tilde\e_i j^5_{\m,i,x}\label{prtr}
\ee
with $\tilde\e_{i_1}=-\tilde\e_{i_2}=1$,
$j_{\m,i,x}=\bar\psi_{i,x}\g_\m\psi_{i,x}$, $ j^5_{\m,i,x}=\bar\psi_{i,x}\g_5\g_\m\psi_{i,x}$
and $\psi_{i,x}=(\psi^-_{i,L,x},\psi^-_{i,R,x})$, $\bar\psi_{i,x}=(\psi^+_{i,L,x},\psi^+_{i,R,x})\g_0$.
Note the chiral nature of the theory, as in the current the fermion with different chirality have different charges.
An example of chiral theory is obtained setting $Q_{i_2}=0$; in such a case one is 
describing $N$ fermions with the same chirality interacting with a gauge field. 
A physically more important example 
is given by the $U(1)$ sector of the Standard Model with no Higgs and massless fermions;
in this case $N=4,$
$i_1=(\n_1,e_1,u_1,d_1)$ are the left handed components and $i_2=(\n_2,e_2,u_2,d_2)$ the right handed of the leptons and quarks.
A formal application of Noether theorem with classical fermions  and bosons says that
the invariance under phase and chiral symmetry,implying the current conservation 
$\partial_\m j^T_\m=0$. If the fermions are quantum (and the bosons classical) the conservation of current is reflected in Ward Identities,
and it turns out
that {\it anomalies} generically break the conservation of $j^T_{\m,i,x}$ unless
\be
\sum_{i_1=1}^N Q_{i_1}^3-\sum_{i_2=1}^N Q_{i_2}^3
=0\label{h}
\ee
In the elecroweak sector
the physical values
$Q_{\n_1}=Q_{e_1}=-1$, $Q_{u_1}=Q_{d_1}=1/3$,
$Q_{\n_2}=0$, $Q_{e_2}=-2$, $Q_{u_2}=4/3$, $Q_{d_2}=-2/3$ verify \pref{h}, if $Q$
are the hypercharges and an index for the three colours of quarks is added. Remarkably the hyperchrges (and therefore the charges)
are constrained to physical values by purely quantum effects.
The question is therefore if in a lattice regularization of \pref{h111}
and considering $A_\m$ a quantum field, the chiral current is conserved under the same condition \pref{h} at a non-perturbative level.

\subsection{The lattice chiral gauge theory}

The lattice chiral gauge theory is defined by its generating function 
\be
e^{\WW(J,J^5,\phi)}= \int P(d A) \int P(d\psi) e^{V(\psi, A, J)+V_c(\psi)+
\BB(J^5, \psi)+(\psi,\phi)}
\label{llll}
\ee
where $A_{\m,x}: \L\to \RRR$, $\L=[0,L]^4\cap a \ZZZ^4$, $L=K a$, $K\in \NNN$
$e_\m$, $\m=0,1,2,3$ an orthonormal basis, $A_{\m,x}=A_{\m, x+L e_\m}$
(periodic boundary conditions) 
and the bosonic integration is
\be
P(dA)={1\over \NN_A} [\prod_{x\in\L} \prod_{\m=0}^3 d A_{\m,x}] e^{-S_G(A)}
\ee
with
\be
S_G=
a^4 \sum_{x}[{1\over 4}F_{\m,\n,x} F_{\m,\n,x}+{M^2\over 2}
 A_{\m,x} A_{\m,x}+(1-\x) (d_\m A_\m)^2]\ee 
is the action of a non-compact lattice $U(1)$ gauge field with a gauge fixing and a mass term,  
$F_{\m,\n}=d_\n A_\m-d_\m A_\n$ and 
$d_\n A_\m=a^{-1}(A_{\m,x+e_\n a}-A_{\m,x})$,
$\NN_A$ is the normalization. The {\it bosonic simple expectation} 
\be
\EE_A(A_{\m_1,x_1}...  A_{\m_n,x_n})=\int P(dA) A_{\m_1,x_1}...  A_{\m_n,x_n}\label{ap}
\ee
is expressed by the Wick rule with covariance
\be
g^A_{\m,\n}(x,y)=
{1\over L^4}\sum_k  { e^{i k (x-y)}\over |\s|^2+M^2}(\d_{\m,\n}+ {\x \bar\s_\m \s_\n\over (1-\x) |\s|^2+M^2})\label{uff} \ee
with $\s_\m(k)=(e^{i k_\m a}-1)a^{-1}$, $k=2\pi n/L$, $n\in \NNN^4$ and $k\in [-\pi/a,\pi/a)^4$.
The {\it bosonic truncated expectation}
\begin{equation}\EE^T_A (F; \cdots; F )=\frac{\partial^n}{\partial \lambda^n}\log\int P(dA)\,  e^{
F(A)}\Big|_{\lambda\equiv 0}\label{dasp}
\end{equation}
is expressed by the Wick rule restricted to the  {\it connected} terms.

We denote by
$\psi^\pm_{i,s,x}$ the Grassmann variables, with $i=1,..,2N$ the {\it particle index}; $s=L,R$ the {\it chiral index}; anti-periodic boundary conditions are imposed and
\be
\{\psi^+_{i, s,x},\psi^+_{i',s',,x'}\}=\{\psi^+_{i, s,x},\psi^-_{i', s',x'}\}=\{\psi^-_{i,s,x},\psi^-_{i',s',x'}\}=0\ee
We define $\psi^\pm_{i,s,x}={1\over L^4}\sum_k e^{\pm i k x}\hat\psi^\pm_{i,s,k}$, with $\hat\psi^\pm_{i,s,k}$ another set of Grassmann variable,
$k=2\pi/L(n+1/2)$, $n\in \NNN^4$ and $k\in [-\pi/a,\pi/a)^4$.
The
fermionic gaussian measure is defined as, $i=1,..,2N$, $s=L,R$
\be
P(d\psi)={1\over\NN_\psi}  [\prod_{i,s,x} d\psi^+_{i,s,x}  d\psi^-_{i,s,x}] 
e^{-S_F}
\ee
where $\NN_\psi$ a normalization and, if $\psi^\pm_{i,x}=(\psi^\pm_{i,L,x},\psi^\pm_{i,R,x})$
%
%
\bea
&&S_F= {1\over 2 a}\sum_{i=1}^{2N} a^4 \sum_x  [\sum_\m  
 (\psi^+_{i,x}\g_0\g_\m   \psi^-_{i, x+e_\m a}-\psi^+_{i,s, x+e_\m a}\g_0\g_\m   \psi^-_{i, x})+\nn\\
&&r ( \psi^+_{i, x}\g_0 \psi^-_{i, x+e_\m a}+\psi^+_{i, x+e_\m a}\g_0 \psi^-_{i, x}
-\psi^+_{i, x}\g_0\psi^-_{i, x}) ]
\eea
We can write therefore
\bea
&&S_F= {1\over 2 a} \sum_{i=1}^{2N}
a^4 \sum_x
[\sum_\m
\sum_{s=L,R}(\psi^+_{i,s, x}\s_\m^s   \psi^-_{i,s, x+e_\m a}-\psi^+_{i,s, x+e_\m a}\s^s_\m \psi^-_{i,s, x})+r ( \psi^+_{i,L, x}\psi^-_{i,R, x+e_\m a}+\nn\\
&&\psi^+_{i,L, x+e_\m a} \psi^-_{i,R, x}
-\psi^+_{i,L, x}\psi^-_{i,R, x} 
 +\psi^+_{i,R, x}\psi^-_{i,L, x+e_\m a}+\psi^+_{i,R, x+e_\m a} \psi^-_{i,L, x}
-\psi^+_{i,R, x}\psi^-_{i,L, x}) ]\label{ssss}
\eea
The {\it fermionic simple expectation} 
\be
\EE_\psi(\psi^{\e_1}_{i_1,x_1}...\psi^{\e_n}_{i_n,x_n})=\int P(d\psi) \psi^{\e_1}_{i_1,x_1}...\psi^{\e_n}_{i_n,x_n}
\ee
is expressed by the anticommutative Wick rule with covariance
\be
g^\psi_i(x,y) =\int P(d\psi) \psi_{i, x} \bar\psi_{i, y}= {1\over L^4}\sum_k e^{i k(x-y)} \hat g^\psi_{i}(k)
\ee
with 
\be
\hat g_{i,k}=
(\sum_{\m}i \g_0\g_\m a^{-1}  \sin (k_\m a)+ a^{-1}\g_0 r \sum_\m (1-\cos k_\m a))^{-1}\label{propp}
\ee
The interaction is
\bea
&&V(A,\psi,J)=V_1(A,\psi,J)+V_2(A,\psi,J)\\
&&V_1(A,\psi,J)=
a^4\sum_{i,s,x}  [O^+_{\m,i,s,x} G_{\m,i,s,x}^+ +
O^-_{\m,i,s,x} G_{\m,i,s}^- ]\nn\\
&&V_2(A,\psi,J)={r\over 2} a^4\sum_{i,x} [ \psi^+_{i,L, x}H_{\m,i,x}^+
\psi^-_{i,R, x+e_\m a}
+\nn\\
&&\psi^+_{i,L, x+e_\m a} H_{\m,i,x}^-\psi^-_{i,R, x}
 +\psi^+_{i,R, x}H_{\m,i,x}^+\psi^-_{i,L, x+e_\m a}+\psi^+_{i,R, x+e_\m a} 
H_{\m,i,x}^-\psi^-_{i,L, x}]
\eea
with
\bea
&&G^{\pm}_{\m,i,s}(x)=a^{-1}  (:e^{\mp i a  Q_i(\l  b_{i,s} A_{\m,x}+ J_{\m,x})}:-1)
\quad H_{\m,i,x}^\pm=
a^{-1}  (e^{\mp i a Q_i J_{\m,x}}-1)
\nn\\
&&O^+_{\m,i,s,x}={1\over 2 }\psi^+_{i,s, x}\s_\m^s   \psi^-_{i,s, x+e_\m a}\quad\quad
O^-_{\m,i,s,x}
=
-{1\over 2 }\psi^+_{i,s, x+e_\m a}\s^s_\m \psi^-_{i,s, x}
\eea
with, if $i_1=1,,N$ and $i_2=N+1,...,2N$
\be b_{i_1,L}=b_{i_2,R}=1;\quad\quad b_{i_1,R}= b_{i_2,L}=0\ee
and $
:e^{\pm i a \l Q_i  A_\m(x)
 }:=e^{\pm i \l Q_i  a A_\m(x)
 } e^{{1\over 2}
(\l Q_i)^2  a^2  g^A_{\m,\m} (0,0) }$.

The mass counterterm is
\be
V_c=  \sum_{i} a^{-1}  \n_i  a^4\sum_x (\psi^+_{i,L,x}\psi^-_{i,R,x}+\psi^+_{i,R,x}\psi^-_{i,L,x})
\ee
Finally the source term is
\be \BB=a^4 \sum_{\m,x} J^5_{\m,x} j^5_{\m,x}
\quad 
j^5_{\m,x}=\sum_{i,s} \tilde \e_i \e_s Q_j Z_{i,s}^5 \psi^+_{x,i,s}\s_\m^s \psi^+_{x,i,s}\nn
\ee
with $\tilde \e_{i_1}=-\tilde \e_{i_2}=1$ and $\e_{L}=-\e_{R}=1$. $\n_i$
and $Z_{i,s}^5 $ are parameters to be fixed by the renormlization conditions, see below.
\vskip.3cm
\noindent
{\bf Remark.} The term proportional to $r$ in $S_F$ \pref{ssss}
is called {\it Wilson term}. If $r=0$ the fermionic propagator \label{propp}
$\hat g_{i,k}$ has, in the $L\to\io$ limit, several poles; this has the effect that the low energy behaviour of the lattice theory
would not correspond to the continuum target theory \pref{h111}; the presence of the Wilson term $r\not=0$
has the effect that only the physical pole $k=0$ is present but the chiral symmetry is broken \cite{NN}.


\subsection{Physical observables}

The fermionic 2-point function is
\be
S^\L_{i,s,s'}(x,y)
={\partial^2 \over \partial \phi^+_{i,s,x}\partial \phi^-_{i,s',y} }
\WW_\L(J,J^5,\phi)
|_0\ee 
and the Fourier transform is 
\be
\hat S^\L_{i,s,s'}(k)=a^4 \sum_x  S^\L_{i,s,s'}(x,0) e^{-i k x}
\ee
The vertex functions are 
\bea
&&\G^\L_{\m,i',s}(z,x,y) =
{\partial^3 \over \partial J_{\m, z}
\partial\phi^+_{i',s,x}\partial \phi^-_{i',s,y} }
\WW(J,J^5,\phi)|_0\nn\\ 
&&\G^{5,\L}_{\m, i' s}(z,x,y) =
{\partial^3 \over \partial J^5_{\m, z}
\partial\phi^+_{i',s,x}\partial \phi^-_{i',s,y} }
\WW(J,J^5,\phi)|_0
\label{h2}
\eea
The Fourier transform is 
\be
\hat \G^\L_{\m,i',s}(k,p) = a^4 \sum_z a^4\sum_y  e^{-i p z-i k y}  \G^\L_{\m,i',s}(z,0,y)
\ee
and similarly is defined $\hat \G^{5,\L}_{\m,i' s}(k,p) $. The three current 
vector $VVV$ and axial $AVV$ correlations are
\be
\Pi^{\L}_{\m,\n, \r}(z,y,x)=
{\partial^3 \WW_\L\over \partial J_{\m,z}
\partial J_{\n,y}\partial J_{\r,x}}|_0;\quad
\Pi^{5,\L}_{\m,\n, \r}(z,y,x)={\partial^3 \WW_\L\over \partial J^5_{\m,z}
\partial J_{\n,y}\partial J_{\r,x}}|_0
\ee
and 
\bea
&&
\hat\Pi^{\L}_{\m,\n, \r}(p_1,p_2)=a^4 \sum_y a^4 \sum_x e^{-i p_1 y-ip_2 x}\Pi^{\L}_{\m,\n, \r}(0,y,x)\nn\\
&&\Pi^{5,\L}_{\m,\n, \r}(p_1,p_2)=a^4 \sum_y a^4 \sum_x e^{-i p_1 y-ip_2 x}\Pi^{5,\L}_{\m,\n, \r}(0,y,x)
\eea
%

\subsection{Ward Identities}
%
%

%
%
%
%
The correlations are connected by relations known as
{\it Ward Identities}. They can be obtained by performing the change of variables
\be
 \label{wis}\psi^\pm_{i,s,x}\to
\psi^\pm_{i,s,x} e^{\pm i Q_i \a_{x} }
\ee
with $\alpha_x$ is a function on $a \ZZZ^4$, with the periodicity of $\L$.
Let $Q(\psi^{+}, \psi^{-})$ be a monomial in the Grassmann variables and 
$Q_{\alpha}(\psi^{+}, \psi^{-})$ be the monomial obtained performing the replacement \pref{wis} in $Q(\psi^{+}, \psi^{-})$.
It holds that
\begin{equation}\label{eq:Jac}
\int [\prod_{i,s,x} d\psi^+_{i,s,x}  d\psi^-_{i,s,x}]
 Q(\psi^{+}, \psi^{-}) = \int [\prod_{i,s,x} d\psi^+_{i,s,x}  d\psi^-_{i,s,x}] Q_{\alpha}(\psi^{+}, \psi^{-})\;.
\end{equation}
as both the left-hand side and the right-hand side of \pref{eq:Jac} are zero unless  the same Grassmann field 
appears once in the monomial, hence
the fields $\psi^+_{i,s,x}$ ,$\psi^-_{i,s,x}$ 
come in pairs and the $\a$ dependence cancels.
By linearity of the Grassmann integration, the property (\ref{eq:Jac}) implies fhe following identity, valid for any function $f$ on the finite Grassmann algebra:
\begin{equation}\label{eq:Jac2}
\int [\prod_{i,s,x} d\psi^+_{i,s,x}  d\psi^-_{i,s,x}]
 f(\psi) = \int [\prod_{i,s,x} d\psi^+_{i,s,x}  d\psi^-_{i,s,x}] f_{\alpha}(\psi)\;,
\end{equation}
with $f_{\alpha}(\psi)$ the function obtained from $f(\psi)$, after the transformation \pref{wis}. 
We apply now \pref{eq:Jac2} to
\pref{llll}; the phase 
in the non-local terms can be exactly compensated by modifying $J$, that is
we get  
\be
W(J,J^5,\phi)=
W(J+d_\m \a ,J^5, e^{i Q \a}
 \phi)\ee
where $J+d_\m \a$ is a shorthand for $J_{\m,x}+d_\m \a_{x}$
and $e^{i Q\a}\phi$ is a shorthand for $e^{\pm i Q_i\a_{x}} \phi_{i,s, x}^{\pm}$; by differentiating we get the Ward Identities (WI)
\bea
&&\sum_\m \s_\m(p)
\hat \Pi^\L_{\m,\n_1,..,\n_n}(p_1,.,p_{n})=0\quad p=p_1+..p_n\nn\\
&&
\sum_\m \s_\m(p) \hat\G^\L_{\m,i,s}(k,p) 
= Q_i
(\hat S^\L_{i,s,s}(k)-\hat S^\L_{i,s,s}(k+p))\label{wia}\\
&&
\sum_\n \s_\n(p_1)
\hat \Pi^{5,\L}_{\m,\n,\r}(p_1,p_2)=\sum_\r \s_\r(p_2)
\hat \Pi^{5,\L}_{\m,\n, \r}(p_1,p_2)=0\nn
\eea

{\bf Remark} The above Ward Identities represent the conservation of the vector part of the current coupled to the gauge field $A_\m$;
in particular the first is the lattice counterpart of $\partial_\m <j_{\m}^{T,V};j_{\n_1}^{T,V};...j_{\n_n}^{T,V}  >_T=0$, see \pref{prtr}.

\subsection{Main result}

Our main result is the following, denoting by $\lim_{L\to\io} \hat S^\L_{i,s,s'}(k)$ and similarly the other correlations.

\begin{theorem} Let us fix $r=1$ and $Ma\ge 1$.
There exists $\l_0,C$ independent on $L,a,M$ such that, 
for $|\l|\le \l_0 (Ma)$, it is possible to find $\n_i$,  $Z_{i,s}^5$ continuous functions in $\l$ such that
\vskip.2cm
\noindent
1) The limits $L\to\io$ of $\hat S^\L_{i,s,s'}(k)$, $\hat \G^\L_{\m,i',s}(k,p) $, $\hat \G^{5,\L}_{\m,i',s}(k,p) $
$\Pi^{\L}_{\m,\n, \r}(p_1,p_2)$,$\Pi^{5,\L}_{\m,\n, \r}(p_1,p_2)$ exist
and $\lim_{k\to 0}
\hat S^\L_{i,s}(k)=\io $ and $\lim_{k,p\to 0}
{\hat \G^{5,\L}_{\m,i',s}(k,p)\over \hat \G^\L_{\m,i',s}(k,p)}=\e_s I$ where $\e_L=-\e_R=1$.
\vskip.2cm
\noindent
2) The AVV correlation verifies
\be
\sum_\m \s_\m(p_1+p_2) \hat\Pi^{5}_{\m,\r,\s}(p_1,p_2)=
\sum_{\m,\n} \e_{\m,\n,\r,\s}
{1\over 2 \pi^2}\s_\m( p^1) \s_\n(p^2) [\sum_{i_1} Q_{i_1}^3-\sum_{i_2} Q_{i_2}^3] +r_{\r,\s}(p_1,p_2)\label{1} 
\ee
with $|r(p_1,p_2)|\le C a^\th \bar p^{2+\th}$, 
$\bar p=\max (|p_1|, |p_2|)$ and $\th=1/2$.
\end{theorem} 
\vskip.3cm
{\bf Remarks}
\begin{enumerate}
\item The correlations are written in the form of expansions which are convergent 
in the limit of infinite volume provided that 
the lattice cut-off is smaller than the boson mass. 
\item
The counterterms $\n_i$
are chosen so that the fermions remain massless in presence of interactions;
the parameters $Z_{i,s}^5$ are fixed so that the charge associated to the vector and axial current are the 
same, a condition present also at a perturbative level \cite{AB}. 
\item Under the
condition $[\sum_{i_1} Q_{i_1}^3-\sum_{i_2} Q_{i_2}^3]=0$ we have
$
\sum_\m \s_\m(p)
\hat \Pi^\L_{\m,\n,\s}(p_1,p_{2})=0$ and 
$\sum_\m \s_\m(p) \hat\Pi^{5,\L}_{\m,\r,\s}(p_1,p_2)=O(a^\th \bar p^{2+\th})$
expressing the conservation of the chiral current in the sense of correlations and up to subdominant terms
for momenta far from the cut-off. The vanishing of the anomaly, obtained up to now only at a purely perturbative level, is proved 
with a finite lattice cut-off, even if the cut-off breaks important symmetries \cite{NN} on which the perturbative cancellation 
were based, like the Lorentz or the chiral one, and excluding non perturbative effects. The anomaly cancellation condition is the same as 
in the continuum case. The lattice regularization plays an essential role; with momentum one a much weaker result holds
\cite{Ma1}. 
\item Anomalies are strongly connected with transport properties in condensed matter \cite{F1}-\cite{F3}
and we use indeed techniques recently developed for the proof of universality properties
in metals to the anomaly cancellation on a lattice \cite{M0}-\cite{BGM1}. Such methods have their roots in the Gallavotti tree expansion 
\cite{Ga}, the Battle-Brydges-Federbush formula \cite{Br}  and the Gawedzki-Kupiainen-Lesniewski
formula \cite{Gaw},  \cite{Le}
(see eg \cite{M5} for an introduction). 
\end{enumerate}

\subsection{Future perspectives}

We have constructed the theory
assuming that $1/a\le (\l_0/|\l|) M$, that is the cut-off is smaller than the boson mass and
we have established \pref{1} for generic values of the coupling.
In this regime after the integration of the $A_\m$ the theory have
scaling dimension
$D=4+n-3 n^\psi/2$ if $n$ is the order and $n_\psi$ the number of fields.
This requires that the "effective coupling"
$\l^2/M^2$ times the energy cut-off must be not too large
so that the expansions are convergent. In order to reach higher 
cut-off one notes that
the boson propagator \pref{uff} is composed by two terms; one which behaves as $O(1/k^2)$ for $k^2>>M^2$ and the other which is 
$O(1)$ for $k^2>>M^2$. If the second term does not contribute the scaling dimension improves and it corresponds to a renormalizable theory
$D=4-3 n^\psi/2-n^A$, so in principle 
one can consider cut-offs higher than $M$ and
up to an exponentially large values $|\l^2 \log a |\le \e_0$. In order to have that the second term does not contribute
full gauge invariance
(broken in our case by the mass and gauge fixing term) is not necessarily required 
but is sufficient the gauge invariance in the external fields, expressed in the form of Ward Identities. It is indeed known
that renormalizability is preserved in QED, at the perturbative level, even if a mass is added to photon, see e.g. \cite{O},\cite{MS}
; if one restricts to gauge invariant observables 
the contribution of the not-decaying term of the propagator is vanishing as consequence of the current conservation.
To get exponentially high cut-off in  $d=4$ QED at a non-perturbative level is technically demanding, as
it would require a simultaneous decomposition in the bosons and fermions, but
the analogous statement can be rigorously proven in 
in $d=2$ vector models \cite{Ma2}.

In the absence of the Wilson term $r=0$ we get the conservation of the chiral current in the form of a WI given by the first of \pref{wia}, if 
$\hat \Pi_{\m,\n_1,...,\n_n}$ is obtained 
replacing 
$J_\m$  in $G^{\pm}_{\m,j,s}$ with
$b_{i,s} \tilde J_{\m,x}$.
As a consequence the averages of 
invariant observables are $\x$ independent. This follows from $\partial_\x \int P(dA) \int [\prod_{i,s,x} d\psi^+_{i,s,x}  d\psi^-_{i,s,x}]  O=0$, with $O(A,\psi)$ invariant;
indeed
\be
\partial_\x \int P(dA)\int [\prod_{i,s,x} d\psi^+_{i,s,x}  d\psi^-_{i,s,x}] 
 O={1\over L^4 }\sum_p 
\partial_\x \hat g^{-1}_{\m,\n}(p) \int P(dA) \hat A_{\m,p} \hat A_{\n,-p}
\int [\prod_{i,s,x} d\psi^+_{i,s,x}  d\psi^-_{i,s,x}]  O
\ee
from which we get, using that $\hat A_{\m,p}=\hat g^A_{\m,\r} {\partial\over \partial A_{\r,-p}}$
\be
\hat g^A_{\m,\r' }(p) \partial_\x (\hat g^A(p))^{-1}_{\m,\n}   \hat g^A_{\n,\r }(p)
{\partial^2\over \partial \hat J_{\r,p}\partial \hat J_{\r',-p}} \int P(dA) \int 
[\prod_{i,s,x} d\psi^+_{i,s,x}  d\psi^-_{i,s,x}] 
O(A+\tilde J,\psi)|_0 \label{ss11}
\ee
By noting that $\partial (\hat g^A)^{-1}=- (\hat g^A)^{-1} \partial_\x \hat g^A( \hat g^A)^{-1}$ 
and   $\partial_\x \hat g^A$ is proportional to $\bar\s_\m \s_\n$, 
by using  $\partial_\a \int P(dA) \int [\prod_{i,s,x} d\psi^+_{i,s,x}  d\psi^-_{i,s,x}]  O(A+d \a,\psi)|_0=0$
then $\partial_\x W$ is vanishing. Therefore if $r=0$ in invariant observables
one can set $\x=0$ and the theory is perturbatively renormalizable.
One expects to be able to reach exponentially high cut-off. 

The Wilson term $r\not=0$, physically necessary to avoid fermion doubling \cite{NN},  breaks the WI and the conservation of chiral current for generic 
values of the charges,
according to \pref{1}. Therefore generically the theory is non-renormalizable at scales greater than $M$ and one cannot expect
in general to be able to reach exponentially high cut-offs.
However choosing the charges so that $[\sum_{i_1} Q_{i_1}^3-\sum_{i_2} Q_{i_2}^3]=0$ the contribution of the non decaying term vanishes up to subdominant terms, making possible in principle to reach exponentially high cut-offs. 
The anomaly cancellation for $1/a\le M$ is therefore a prerequisite
for reaching higher cut-offs.
In the case of the $U(1)$ sector of the Standard Model, one has also to introduce an Higgs boson to generate the fermion mass;
one can distinguish a region higher than the boson mass generated by the Higgs, where the second term
of the boson propagator does not contribute due to the anomaly cancellation and the WI; 
and a lower one, when 
the infinite volume limit can be taken using the infrared freedom of QED and the massive nature of weak forces. Further challenging problems
arise considering the anomaly associated to the $SU(2)$ sector.

\section{Proof of Theorem 1.1} 

In the following we denote by $C$ or by $C_1,C_2..$ generic $\l,L,a$-independent constants.
We integrate the bosonic variables $A_\m$ in \pref{llll}, obtaining
\be 
V_F(\psi,J)= \log \int P(dA) e^{V_1(\psi, A, J)}
\ee
%
%
where, by \pref{dasp}
\bea
&&V_F(\psi,J)=a^4\sum_x \sum_{i,\e=\pm}  a^{-1} 
(e^{-i a \e  Q_{i} J_{\m,x}}-1)  O^\e_{\m,s,i}+
\sum_{n=2}^\io\
a^{4 n} \sum_{x_1,...,x_n}  \sum_{\underline\e,\underline i\atop \underline \m,\underline s}\nn\\
&&[\prod_{j=1}^n O^{\e_j}_{i_j,\m_j,s_j,x_j}e^{i\e_j a Q_{i_j}  J_{\m_j,x_j}}] {1\over n!} a^{-n}\EE^T_A( : e^{i\e_1 b_{i_1,s_1}\l a Q_{i_1}  A_{\m_1,x_1} }:  ;...; 
:e^{i \e_n b_{i_n,s_n}  \l a Q_{i_n} A_{\m_n, x_n}}:)\nn
\eea
which can be rewritten as, if $\underline x=x_1,..,x_n$, $\underline i=i_1,...,i_n$,
$\underline \m=\m_1,...,\m_n$,$\underline m=m_1,..,m_n$
\bea
&&V_F(\psi,J)=a^4\sum_x \sum_{i,\e=\pm}  a^{-1} 
(e^{-i a \e Q_{i} J_{\m,x}}-1)  O^\e_{\m,s,i}+\label{ap1}\\
&&\sum_{n=2}^\io \sum_{m=0}^\io a^{4n} \sum_{x_1,...,x_n}
\sum_{\underline\e,\underline i,\underline \m,\underline s,\underline m \atop \sum_j m_j=m}
{1\over n!} 
 [\prod_{j=1}^n O^{\e_j}_{i_j,\m_j,s_j,x_j} (J_{\m_j,x_j})^{m_j}
]
H_{n,m}(\underline x,\underline\e,\underline i,\underline \m,\underline s,\underline m)\nn
\eea
with
\be
H_{n,m}(\underline x,\underline\e,\underline i,\underline \m,\underline s,\underline m)= {a^{-n}
\over n!}
{(i a \e_{j}  Q_{i_j})^{m_{j}} \over m_{j}!}]
\EE^T_A( : e^{i\e_1 b_{i_1,s_1}\l Q_{i_1}  A_{\m_1,x_1} }:  ;...; 
:e^{i \e_n b_{i_n,s_n}  \l a Q_{i_n} A_{\m_n,x_n} }
)\label{ap2a}
\ee
and
$
||H_{n,m}||=L^{-4} a^{4 n}\sup_{\underline\e,\underline i,\underline \m,\underline s,\underline m \atop \sum_j m_j=m}
\sum_{x_1,...,x_n} |H_{n,m}|$.
\begin{lemma} The kernels in \pref{ap1} the following bound, for $n\ge 2$, $m\le 3$ and uniformly in $L$
\be
||H_{n,m}||
\le C^{n} a^{-(4-3 n-m)N} (|\l|/(M a))^{2(n-1)} \label{sh}
\ee
\end{lemma}
\noindent
{\bf Proof of Lemma 2.1}
We write the truncated expectations in \pref{ap2a}
by the Battle-Brydges-Federbush formula, see e.g. Theorem 3.1 in \cite{Br} (for completeness a sketch of the proof is in Appendix 1), $n\ge 2$
\be
\EE^T_A(  e^{i\e_1 b_{i_1,s_1}\l Q_{i_1}  A_{\m_1,x_1} } ;...; 
e^{i \e_n b_{i_n,s_n}  \l a Q_{i_n} A_{\m_n,x_n} })=
\sum_{T\in \boldsymbol T_n}\prod_{\{j,j'\}\in T} \tilde g^A_{\mu_j\mu_{j'} }(x_j, x_{j'})\int dp_T(\underline t) e^{-V(X;\underline t)},\label{ep}
\ee
where: $X=((x_1,\e_1,i_1,\m_1,s_1,m_1);..;
(x_n,\e_n,i_n,\m_n,s_n,m_n))$,
${\bf T}_n$ is the set of connected tree graphs on $\{1,2,\ldots,n\}$, the product $\prod_{\{i,j\}\in T}$ runs over the edges of the tree graph $T$,
\be
\tilde g^A_{\mu_j\mu_{j'}}(x_{j} , x_{j'})=\l^2 a^2 \e_j b_{i_j,s_j} Q_{i_j} \e_{j'}  b_{i_{j'},s_{j'}} Q_{i_{j'}}  g^A_{\mu_j\mu_{j'}}(x_{j} , x_{j'}),
\ee
$V(X;\underline t)$ is obtained by taking a sequence of convex linear combinations, with parameters $\underline t$, of the energies $V(Y)$ of suitable subsets $Y\subseteq X$, defined as
\be
V(Y)=\sum_{j,j'\in Y}\l^2 \e_j\e_{j'} b_{i_j,s_j} Q_{i_j} b_{i_{j'},s_{j'}} Q_{i_{j'}} 
a^2 g^A_{\m_j,\m_{j'} }(x_j, x_{j'}) =
\EE_A( [\sum_{j\in Y}  \l b_{i_j,s_j} Q_{i_j}  a \e_j A_{\m_j}(x_j)]^2)
\ee
and $dp_T(\underline t)$ is a probability measure, whose explicit form is recalled in the Appendix.
We use the bounds
\be
|g^A_{\m,\n}(x,y)|_1=a^4 \sum_x |g^A_{\m,\n}(x,y)|\le C M^{-2}\quad\quad |g^A_{\m,\m} (x,y)|\le C a^{-2}
\label{ap3}  
\ee
so that
\be
||H_{n,m}||\le C_1^n
L^{-4} \sup_{\underline\e,\underline i,\underline \m,\underline s,\underline m \atop \sum_j m_j=m} a^{4 n}\sum_{x_1,...,x_n}
{a^{-n+m}\over n!}\sum_{T\in \boldsymbol T_n}\prod_{\{j,j'\}\in T} |\tilde g^A_{\mu_j\mu_{j'} }(x_j, x_{j'})|\int dp_T(\underline t) e^{-V(X;\underline t)}|\label{ep1}
\ee
Moreover
$V(Y)$ is stable, that is
\be
V(Y)=\EE( [\sum_{j\in Y}  \l b_{i_j,s_j} Q_{i_j}  a \e_j A_{\m_j}(x_j)]^2)\ge 0
\ee
hence $
V(X;\underline t)\ge 0$ and $e^{-V(X;\underline t)}\le 1$ so that 
$\int dp_T(\underline t) e^{-V(X;\underline t)}<1$ therefore
\be
||H_{n,m}||\le C_2^n
{a^{-n+m}\over n!}\sum_{T\in \boldsymbol T_n}\prod_{\{j,j'\}\in T} 
a^2 | g^A_{\mu_j\mu_{j'} }(x_j, x_{j'})|_1\le C_3^n
{a^{-n+m}\over n!}\sum_{T\in \boldsymbol T_n} (a M^{-1})^{2(n-1)} 
\label{ep2}
\ee
and finally using that $\sum_{T\in \boldsymbol T_n} 1\le C_4^n n!$ by Cayley' formula
\cite{Ca}
we finally get
\be
||H_{n,m}||\le C_5^n
a^{-n+m} (a M^{-1})^{2(n-1)}=C_5^{n} a^{-(4-3 n-m)N} (|\l|/(M a))^{2(n-1)} 
\label{ep2}
\ee
\qed
\vskip.3cm
After the integration of $A_\m$ the generating function can be written 
as a Grassmann integral
\be
e^{\WW(J,J^5,\phi)}=\int P(d\psi) e^{V^{(N+1)}(\psi, J, J^5,\phi)}
\ee
with
\be
V^{(N+1)}(\psi, J, J^5,\phi)
=V_F(\psi,J)+V_2(\psi, J)+V_c(\psi)+
\BB(J^5, \psi)+(\psi,\phi)
\label{llll1}
\ee
The fermionic propagator is massless, that is it has a power law decay at large distances and this requires
a multiscale analysis based on Wilson Renormalization Group.

We introduce parameters $\g>1$ and $N\in \NNN$ such that\footnote{Any choice for $\g^N$ ensuring that in the support of $1-\tilde f$
does not include the doubled poles, that is the poles of $\hat g(k)$ with $r=0$ different from $k=0$, could be done.}
 $\g^N\equiv \pi/(16 a)$
; moreover we introduce $\tilde f(t); \RRR^+\to \RRR $ a $C^\io$ non-decreasing function $=0$ for $0\le t\le \g^{N-1}$ and $=1$ for $t\ge \g^N$;
we define also $\chi_{N}(t)=1-\tilde f(t)$ which is therefore non-vanishing for $t\le \g^N$. We introduce the propagator
\be
g^{(N+1)}_i(x,y) = {1\over L^4}\sum_k e^{i k(x-y)} \tilde f(|k|_T)  \hat g^\psi_{i}(k)
\ee
with $|k-k'|_T$  the distance
on  the 4-dimensional torus $[-\pi/a,\pi/a)^4$. Therefore, for any $K\in \NNN$ we have
\be
|g^{(N+1)}_i(x,y)|\le \g^{3(N+1)} {C_K \over 1+( \g^{N+1}  |x-y|_{\tilde T})^K}\label{lap}
\ee
where $|x-y|_{\tilde T}$ is the distance on the $[-L,L)^4$ torus. The above bound is derived by (discrete) integration by parts, see e.g. \S 3.3 of \cite{M5},
using that\footnote{The bound \pref{lap} follows from the presence of the Wilson term; if $r=0$ a power law is found due to the presence of poles in the support of $\tilde f(k)$.}
 in the support of  $\tilde f(|k|_T)$ one has  $\sum_\m (1-\cos k_\m a)^2/a^2\ge C/a^2$ and
the volume of the support of $\tilde f$ is $\le C /a^4$. 

We can write therefore, using the addition property of gaussian
Grassmann integrals, see e.g. \S 2.4 of \cite{M5}
\bea
&&e^{\WW(J,J^5,\phi)}=\int P(d\psi^{(\le N)})\int P(d\psi^{(N+1)})
 e^{V^{(N+1)}(\psi^{(\le N)}+\psi^{(N+1)}, J, J^5,\phi)
}=\nn\\
&&\int P(d\psi^{(\le N)})
e^{V^{(N)} (\psi^{(\le N)},J,J^5,\phi)}
\label{llll2}
\eea
where 
\bea
&&V^{(N)}(\psi^{(\le N)},J,J^5,\phi)=\sum_{n=1}^\io {1\over n!}\nn\\
&&\EE^T_{N+1}(V^{(N+1)}(\psi^{(\le N)}+\psi^{(N+1)},J,J^5,\phi);...; V^{(N+1)}(\psi^{(\le N)}+\psi^{(N+1)},J,J^5,\phi)  )\label{gigi}
\eea
and $\EE^T_{N+1}$ is the truncated expectation with respect to the integration $P(d\psi^{(N+1)} )$.

Using
the linearity of the truncated expectations, one gets, if $\underline\g=\underline \e,\underline s,\underline i,\underline\m,\underline\b$
\be
V^{(N)}(\psi^{(\le N)}
,J,J^5,\phi)=a^{4 (l_a+l_b+m)}\sum_{\underline x,\underline y,\underline z,\underline\g}
W^{(N)}_{l_a,l_b,m}(\underline x,\underline y,\underline z,\underline \g)
[ \prod_{j=1}^{l_a} \psi^{\le N, \e_j}_{x_j, i_j, s_j}]
[ \prod_{j=1}^{l_b} \phi^{\e_j}_{y_j, i_j, s_j}]
[\prod_{j=1}^m J^{\b_j}_{\m_j,z_j}]\label{ap222} 
\ee
with  $\e=\pm$, 
and $J^{\b}_{x_j}$  is $J_{x_j}$ or $J^5_{x_j}$ for $\b=(0,1)$. 
Note that the $W^{(N)}$ are a series in the kernels $H_{n,m}$. 
In the $l_b=0$ case (the presence of $\phi$ is briefly discussed in the Appendix 1)
calling
$W^{(N)}_{l_a,0,m}\equiv W^{(N)}_{l_a,m}$ we define
$
||W^{(N)}_{l,m}||=L^{-4} \sup_{\underline \g}
a^{4 l+4 m}\sum_{\underline x,\underline z}
|W^{(N)}_{l,m}(\underline x,\underline z,\underline \g)|$.
\begin{lemma} The kernels in \pref{llll2} verify, for $|\l|\le \l_0 (M a)$,  $|\n_i|\le C (|\l|/(Ma))^2$,
$\l_0,C,C_1$ independent on $a,L, N$, $l\le 2, m\le 3$, $|d |$ is the distance between any coordinate in $\underline x, \underline z$ 
\be
|| d^s W^{(N)}_{l,m}||
\le C_1 \g^{D N} \label{hhh}
\ee
with $D=4-3 l/2 -m-s$.
\end{lemma}
\vskip.3cm
{\it Proof of Lemma 2.2}. We rewrite $V^{(N+1)}$ \pref{llll1}
in a more compact way as
\be
V^{(N+1)}=\sum_{P}\tilde\psi^{(\le N+1)}(P)\tilde J(P) 
W^{(N+1)}(P)
\label{ap2g} 
\ee
with $P$ set of field labels and
$
\tilde\psi(P)=\prod_{f\in P} \psi^{(\le N)\e(f)}_{i(f),s(f),x(f)}$, $\tilde J(P) =\prod_{f\in P} J^{\b(f)}_{x(f)}$.
%
We get
therefore, inserting \pref{ap2g} in \pref{gigi}
%
%
if $P=Q_1\cup Q_2...\cup Q_n$
\bea
&&V^{(N)}(\psi^{\le N},J,J^5)=
\sum_n  {1\over n!}\sum_P \sum_{P_1,...P_n\atop Q_1,...Q_n} \tilde\psi^{(\le N)}(P) \nn\\
&&\EE^T_{N+1}(
\tilde\psi^{(N+1)}(P_{1}/Q_{1});
...;\tilde\psi^{(N+1)}(P_{n}/Q_n)) [\prod_{i=1}^n
W^{(N+1)}(P_i)\tilde J(P_i)]\label{60}
\eea
We use the  Gawedzki-Kupiainen-Lesniewski  \cite{Gaw}, \cite{Le} (a sketch of the proof is in App.1; see also 
(see e.g. \S A.3 of \cite{GM}, 
 \S 2 of \cite{M5} or App D of \cite{GMR})
\be
\EE^T_{N+1}(\tilde\psi^{(N+1)}(P_1);
...;\tilde\psi^{(N+1)}(P_s))= \sum_{T\in \TT_n}\prod_{\{i,j\}\in T} g^{(N+1)}(x_i,y_j)
\int dP_{T}(\underline t) \det
G^{N+1,T}(\underline t)\label{xx}\ee
where $\TT_n$ denotes the set of all the ‘spanning trees’ on $x_{P_1},...,x_{P_s}$,
that is a set of lines which becomes a tree graph on $\{1,2,\ldots,s\}$
if one contracts in a point all the point in $x_P=\cup_{f\in P} x(f)$, the product $\prod_{\{i,j\}\in T}$ runs over the unordered edges of the $T$,
$\underline t=\{t_{i,i'}\in [0,1],
1\le i,i' \le s\}$, $dP_{T}(\underline t)$ is a probability measure (whose form is specified in the Appendix) with support on a set of $\underline t$ such that
$t_{i,i'}=u_i\cdot u_{i'}$
for some family of vectors $u_i\in \RRR^s$ of unit norm and
$G^{N+1,T}(\underline t)$ is a $(n-s+1)\times (n-s+1)$ matrix, whose elements
are given by
$G^{N+1,T}_{ij,i'j'}=t_{i,i'} 
g^{(N+1)}(x_{ij},y_{i'j'})$ such that 
if =$<u_i\otimes A^{(N+1)}_{x(f^-_{ij})
},
u_{i'}\otimes B^{(N+1)}_{x(f^+_{i'j'})}>$ then the matrix element can be written as a scalar product 
\be
G^{N+1,T}_{ij,i'j'}
=<u_i\otimes A^{(N+1)}_{x(f^-_{ij})
},
u_{i'}\otimes B^{(N+1)}_{x(f^+_{i'j'})}>=(u_i\cdot u_{i'}) {1\over L^d}\sum_k
\bar A^{(N+1)}_{x(f^-_{ij}),k)} B^{(N+1)}_{x(f^+_{i'j'}),k)}
\ee
with $A^{(N+1)}_{x(f^-_{ij},k)}=e^{i k x(f^-_{ij})}
\sqrt{f^N(k) \bar g^ {N}(k)}$ and $B^{(N+1)}_{x(f^+_{i'j'}),k)}=e^{i k x(f^+_{i'j'}))}
\sqrt{f^N(k) g^ {N}(k)}$.

The determinants are bounded by the {\it Gram-Hadamard inequality}, see e.g. \S 2 of \cite{M5}, stating
that, if $M$ is a square matrix with elements $M_{ij}$ of the form
$M_{ij}=<A_i,B_j>$, where $A_i$, $B_j$ are vectors in a Hilbert space with
scalar product $<\cdot,\cdot>$, then
$
|\det M|\le \prod_i <A_i,A_i>^{1\over 2}  <B_j,B_j>^{1\over 2} $. Therefore 
\be
|\det G^{N+1,T}|\le 
C_1^{\sum_i |P_i|} \g^{3N \sum_ i [|P_i|-(n-1))/2] }) 
\ee
We get, setting 
$|P_i|\equiv n_i$, $|Q_i|\equiv l_i$, $0\le l_i\le n_i$,$\sum_i m_i=m$, $l=\sum_i l_i$ 
and $| |x|^s g^{(N+1)}(x)|_1\le C_2 \g^{- N-s}$
\bea
&&||W^{(N)}_{l,m}||\le \sum_{n=1}^\io  C_3^n {1 \over n!}\sum_{n_1,...n_n}
\sum_{l_1,..l_n\atop \sum_i l_i=l}
[\prod_{i=1}^n {n_i!\over l_i! (n_i-l_i)!} ]\nn\\
&&\sum_T  C_2^s \g^{-N ((n-1)+s)  }
C_1^{\sum_i n_i} \g^{3N [\sum_ i (n_i-l_i)/2-(n-1)] }) 
 [\prod_{i=1}^n
||W^{N+1}(P_i)||]
\eea
and $\sum_{T\in \TT_n}\le n! C_4^{\sum_i n_i}$, see e,g, lemma A3.3 of \cite{GM}, Lemma 2.4 of \cite{M5}  or Lemma 
D.4 of \cite{M5}, so that
\bea
&&||W^{(N)}_{l,m}||\le \sum_{n=1}^\io  C_6^{n} \sum_{n_1,...n_n}
\sum_{l_1,..l_n\atop \sum_i l_i=l} 
[\prod_{i=1}^n (C_5)^{\sum_i n_i}  {n_i!\over l_i! (n_i-l_i!)} ]\nn\\
&&\g^{-4 N (n-1)} \g^{-Ns}  \g^{3 N   (\sum_ i  n_i -l_i)/2 }
[\prod_{i=1}^n
\g^{N(4-3 n_i/2-m_i)}]
[ \prod_{i}  (|\l|/Ma)^{\max (2(n_i/2-1),1-m_i)}]]
\nn  
\eea
Note that $\sum_{l\le n} 
[(C_5)^{n-l} C_5^l  {n!\over l! (n-l)!} ]= 2C_5$ we get
\be
||W^{(N)}_{l,m}||\le \sum_{n=1}^\io  C_8^{n} \g^{N(4-3 l/2-m-s)}]
\prod_{i}[ \sum_{n_i} C_7^{n_i} (|\l|/Ma)^{\max (2(n_i/2-1),1-m_i)}]]
\ee
As $\sum_i m_i=m\le 3$ the sum over $n_i$ is bounded
by  
\be
\prod_{i}[ \sum_{n_i} C_7^{n_i} (|\l|/Ma)^{\max (2(n_i/2-1),1-m_i)}]]\le C_8^n (|\l|/Ma)^{2\max (n-3,1)}
\ee
so that for $\l$ small enough
\be
||W^{(N)}_{l,m}||\le \g^{N  (4-3l-m )}
C_9 (1+\sum_{n=4}^\io C_8^n  (|\l|/Ma)^{2(n-3)}]
\le C_1 \g^{N  (4-3l/2 -m )} 
\ee
\qed
\vskip.3cm
In order integrate 
$\int P(d\psi^{(\le N)})
e^{V^{(N)} (\psi^{(\le N)},J,J^5)}$ \pref{llll2} we need to take into account 
the presence of terms with positive or negative scaling dimension
$D=4-3 l/2-m$, as can be read from \pref{hhh}. 

In order to do that we extract
from $V^{(N)}$ the terms with non negative dimension. This is done 
defining an $\LL$ (localization) linear operation acting on the kernels of $\hat W^N_{l,m}$
(the Fourier transform of $W^N_{l,m}$ in\pref{llll2}) in the following way; $\LL \hat W^N_{l,m}(\underline k)=W^N_{l,m}(\underline k)$
for $(n,m)\not =(2,0), (2,1)$ and
\be
\LL \hat W^N_{2,0}(k)=\hat W^N_{2,0}(0)+{\sin k_\m a\over a}\partial_\m \hat W^N_{2,0}(0)\quad\quad
\LL \hat W^N_{2,1}(k,k+p)=\hat W^N_{2,1,\m}(0,0)\label{loc}
\ee
We write therefore
\be
e^{\WW(J,J^5,0)}=
\int P(d\psi^{(\le N)})
e^{\LL V^{(N)} (\psi^{(\le N)},J,J^5)+\RR V^{(N)} (\psi^{(\le N)},J,J^5)}
\ee
with $\RR=1-\LL$ (renormalization) and $\RR V^{(N)}$
is equal to \pref{ap222} with
$W^{(N)}_{l,m}$ replaced by $\RR W^{(N)}_{l,m}$; the $\RR$
operation produce an improvement in the bound, see
eg \S 4.2 of \cite{M5}; for instance
$\RR \hat W^N_{2,0}(k)$ will admit, by interpolation, a bound similar to the one for 
$\hat W^N_{2,0}(k)$ times a factor $O(\g^{-2 N})$ due to the derivatives and an extra 
$O(\g^{2 h})$, with $h$ the scale associated to the external fields due to the $k^2$. Hence the $\RR$ operation produces on such terms an improvement $O(\g^{2(h- N)})$. In coordinate space, the action consists in producing a derivative in the external field and a "zero", that is the difference of two coordinates, see e.g. \S 3 of \cite{BM}.

Using 
symmetry considerations, see the Appendix 3, we get
\bea
&&\LL\VV^{(N)}(\psi,J,J^5)
=a^4 \sum_x \sum_{i,s}  [n_{N,s,i}  \g^{-N}
(\psi^+_{i,L,x}\psi^-_{i,R,x}+\psi^+_{i,R,x}\psi^-_{i,L,x})+\nn\\
&&
z_{N,i,s}
\s_\m^s\psi^+_{i,s,x}\tilde\partial_\m\psi^+_{i,s,x}\nn+
\tilde Z_{i,s,N}^{J}
J_{\m,x} \psi^+_{i,s,x}\s_\m^s \psi^-_{i,s,x}+\e_s \tilde\e_i \tilde Z_{i,s,N}^{5} J^5_{\m,x}
\psi^+_{i,s,x}\s_\m^s \psi^-_{i,s,x}]\label{ess11a}
\eea
with $\e_L=-\e_R=1$, $\tilde\e_{i_1}=-\tilde\e_{i_2}=1$, 
$n_{N,s,i}  \g^{-N}=\hat W^N_{2,0}(0)$, 
$z_{N, s,i} =\partial_\m \hat W^N_{2,0}(0)$ and 
$\tilde Z_{i,s,N}^{J}=\hat W^N_{2,1}(0,0)$, 
$\tilde Z_{i,s,N}^{5}=\hat W^N_{2,1}(0,0)$ respectively with $J$ and $J^5$.

It is possible to include the marginal quadratic terms in the fermionic gaussian integration in the following way
\be
P(d\psi^{(\le N)})
e^{\sum_{i,s}  z_{h,i,s}Z_{h,s,i} a^4 \sum_x  
\s_\m^s\psi^+_{i,s,x}\tilde\partial_\m\psi^+_{i,s,x}}\equiv P_{Z_N} (d\psi^{(\le N)})
\ee
where $\tilde\partial$ is the discrete derivative and 
\be
\hat g_{i}^{(\le N)}(k)=\chi_N(k)
(\sum_{\m} \g_0\tilde\g^N_\m a^{-1}  i \sin (k_\m a)+  a^{-1}\hat\g^N_0 \sum_\m (1-\cos k_\m a))^{-1}\label{sapa}
\ee
\be
\tilde \g^N_0= \begin{pmatrix} 0& Z_{N,L,i}(k) I  \\  Z_{N,R,i}(k)I & 0  \end{pmatrix}\quad \tilde \g^h_j= \begin{pmatrix} 0& i Z_{N,L,i}(k) \s_j \\-i Z_{N,R,i}(k)\s_j& 0 \end{pmatrix}
\ee
with $Z_{N,s,i}(k)=1+\chi^{-1}_N(k) z_{N,s,i}$, and we set $Z_{N,s,i}\equiv 1+z_{N,s,i}$.
We can write therefore 
\be
e^{\WW(J,J^5,0)}=\int
P_{Z_N}(d\psi^{(\le N)})
e^{\tilde \LL V^{(N)} ( \sqrt{Z_N} \psi^{(\le N)},J,J^5)+\RR V^{(N)} (\sqrt{Z_N}\psi^{(\le N)},J,J^5)}
\ee
where we have rescaled the fields writing
\bea
&&\tilde \LL\VV^{(N)}(\sqrt{Z_N}\psi,J,J^5)
=a^4 \sum_x \sum_{i,s}  [\n_{N,s,i}\g^{-N}
\sqrt {Z_{N,L,i}Z_{N,R,i}}  
(\psi^+_{i,L,x}\psi^-_{i,R,x}+\psi^+_{i,R,x}\psi^-_{i,L,x})+\nn\\
&&+
Z_{i,s,N}^{J}
J_{\m,x} \psi^+_{i,s,x}\s_\m^s \psi^-_{i',s,x}+\e_s \tilde\e_i 
Z_{i,s,N}^{5} J^5_{\m,x}
\psi^+_{i,s,x}\s_\m^s \psi^-_{i,s,x}]\label{ess11a}
\eea
with $\n_{N,s,i}
\sqrt {Z_{N,L,i}Z_{N,R,i}} =n_{N,s,i}$ and 
$\tilde Z_{i,s,N}^{J}/Z_{i,s,N}= Z_{i,s,N}^{J}$, $\tilde Z_{i,s,N}^{5}/Z_{i,s,N}= Z_{i,s,N}$.

We choose $\chi_{N}(t)\equiv \chi_0(\g^{-N} t )$ with 
$\chi_0(t); \RRR^+\to \RRR $ a $C^\io$ non-increasing function $=1$ for $0\le t\le \g^{-1}$ and $=0$ for $t\ge 1$; 
and we write
\be
\chi_N(t)=\sum_{h=-\io}^N f_h(t)\quad\quad f_h(t)=\chi_0(\g^{-h} t )-\chi_0(\g^{-h+1} t )
\ee
with $f_h(t)$ with support in $\g^{h-1}\le t \le \g^{h+1}$. We can write $\chi_{N}(t)=\chi_{N-1}(t)+ f_N(t)$ and
\be
\hat g_{i}^{(\le N)}(k)=\hat g_{i}^{(\le N-1)}(k)+\hat g_{i}^{(N)}(k)
\ee
with 
$\hat g_{i}^{(N)}(k)$ given by \pref{sapa} with $\chi_N(k)$ replaced by 
$f_N(k)$ and $Z_{i,s,N}(k)$ replaced $Z_{i,s,N}$. We write therefore
\bea
&&e^{\WW(J,J^5,0)}=
\int P_{Z_N}(d\psi^{(\le N-1)})\int P_{Z_N}(d\psi^{(N)})
e^{\tilde \LL V^{(N)} ( \sqrt{Z_N}\psi^{(\le N)},J,J^5)+\RR V^{(N)} (\sqrt{Z_N}
\psi^{(\le N)},J,J^5)}=\nn\\
&&\int P_{Z_N}(d\psi^{(\le N-1)})
e^{
V^{(N-1)} ( \sqrt{Z_N} \psi^{(\le N-1)},J,J^5)}
\eea
where
\be
V^{(N-1)}=\sum_n {1\over n!}\EE^T_N(\tilde \LL V^{(N)}+\RR  V^{(N)},
...;\tilde \LL V^{(N)}+\RR V^{(N)}
)\label{60a} 
\ee
with $\VV^N$ given by \pref{60}; a graphical representation is in fig.2. Using more
compact notation
\be
V^{(N-1)}=\sum_{P,\tilde P}\tilde\psi^{(\le N-1)}(P)\tilde J(P) 
W^{(N-1)}(P)
\label{ap2ga} 
\ee
By using the linearity of the truncated expectations and expressing
$\RR V^N$ by \pref{gigi} we can write, calling $\EE^T(V;...;V)=\EE^T(V;n)$
\pref{60a} as, see Fig. 3
\be
V^{(N-1)}=\sum_n {1\over n!}\EE^T_N(\tilde \LL V^{(N)}+
\sum_m {1\over m!}\RR \EE^T_ {N+1}(V^{(N+1)};m);n)\label{60b} 
\ee
From \pref{60b} we see that  $W^{(N-1)}$ is a function of $W^{(N+1)},\n_N, Z_N Z^J_N,Z^5_N$.

The procedure can be iterated in a similar way writing
\be
P_{Z_N}(d\psi^{(\le N-1)}) =P_{Z_N-1}(d\psi^{(\le N-2)})P_{Z_N}(d\psi^{(N-1)}) 
\ee
and $V^{(N-1)}=\LL V^{(N-1)}+\RR V^{(N-1)}$ with $\LL$ acting on the kernels $W^{(N-1)}$
as \pref{loc}, so that, after modifying the wave function renormalization and rescaling, we get to
\be
\int P_{Z_{N-1}}(d\psi^{(\le N-2)})\int P_{Z_{N-1}}(d\psi^{(N-1)})
e^{\tilde \LL V^{(N-1)} ( \sqrt{Z_{N-1}} \psi^{(\le N-1)},J,J^5)+
\RR V^{(N-1)} (\sqrt{Z_{N-1}}\psi^{(\le N-1)},J,J^5)}
\ee
Therefore, after integrating in the same way $\psi^{(N-1)},\psi^{(N-2)},...,\psi^{(h+1)}$
\be
e^{\WW(J,J^5,0)}=\int P_{Z_h}(d\psi^{(\le h)})
e^{
V^{(h)} ( \sqrt{Z_h} \psi^{(\le h)},J,J^5)}\label{dds1}
\ee
with $P_{Z_h}(d\psi^{(\le h)})$ with propagator
\be
\hat g_{i}^{(\le h)}(k)=\chi_{h}(k)
(\sum_{\m} \g_0\tilde\g^{h}_\m a^{-1}  i \sin (k_\m a)+  
a^{-1}\hat\g^{h}_0 \sum_\m (1-\cos k_\m a))^{-1}\label{sapa1}
\ee
\be
\tilde \g^{h}_0= \begin{pmatrix} 0& Z_{h,L,i}(k) I  \\  Z_{h,R,i}(k)I & 0  \end{pmatrix}\quad \tilde \g^h_j= \begin{pmatrix} 0& i Z_{h,L,i}(k) \s_j \\-i Z_{h,R,i}(k)\s_j& 0 \end{pmatrix}
\ee
and 
\bea
&&\tilde \LL\VV^{(h)}(\sqrt{Z_h}\psi,J,J^5)
=a^4 \sum_x \sum_{i,s}  [\n_{h,s}\g^{h}
\sqrt {Z_{h,L,i}Z_{h,R,i}}  
(\psi^+_{i,L,x}\psi^-_{i,R,x}+\psi^+_{i,R,x}\psi^-_{i,L,x})+\nn\\
&&+
Z_{i,s,h}^{J}
J_{\m,x} \psi^+_{i,s,x}\s_\m^s \psi^-_{i',s,x}+\e_s \tilde\e_i 
Z_{i,s,h}^{5} J^5_{\m,x}
\psi^+_{i,s,x}\s_\m^s \psi^-_{i,s,x}]\label{ess11a}
\eea
and finally, if $\underline \g=\underline \a,\underline s,
\underline i,\underline\m,\underline\b$
\be
V^{(h-1)} (\sqrt{Z_h} \psi^{(\le h-1)},J,J^5)
=\sum_{l,m}  a^{4 l+4m}\sum_{\underline x,\underline z}
\sum_{\underline \g}
W^{(h-1)}_{l,m}(\underline x,\underline z,\underline \g)
[ \prod_{j=1}^{l}  \psi^{\le h-1, \e_j}_{x_j, i_j, s_j}]
[\prod_{j=1}^m J^{\b_j}_{\m_j,z_j}]\label{ap222} 
\ee
and
\be
||W^{h-1)}_{l,m}||=L^{-4} \sup_{\underline \g }
a^{4 l+4 m}\sum_{\underline x,\underline z}
|W^{(h-1)}_{l,m}(\underline x,\underline z)|
\ee
The $\n_{k,i}$ is a relevant running coupling constant representing the the renormalization of the mass of the fermion of type $i$; 
$\ZZ_{k,i,s}=(Z_{k,i,s},Z^J_{k,i,s},Z^5_{k,i,s})$ are the marginal couplings
and represent respectively the wave function renormalization of the fermion 
of type $i$ and chirality $s$, and the renormalization of the current
and of the axial current.
By construction  $W^{(h-1)}$ is a function of the kernels $W^{(N+1)}$
in $V^{N+1}$  and of the running coupling constants
$\n_N, \ZZ_N,...,\n_h,\ZZ_h$; moreover, the running coupling constants
verify recursive equations of the form
\be
\n_{h-1,i}=\g\n_{h,i}+\b^h_{\n,i}(\n_N,...,\n_h, W^{(N+1)})\quad
\ZZ_{h-1,i,s}=\ZZ_{h,i,s}+\b^h_{\n,i,s}(\n_N,\ZZ_N...,\n_h,\ZZ_h, W^{(N+1)})
\ee
As should be clear from the previous pictures, 
the $W^h$ and the $\b^h$ can be conveniently represented
in terms as a sum of labeled trees, called Gallavotti trees, 
%
, see Fig.4, defined in the following way (for details see e.g.   \S 3 of \cite{M5}) .
\insertplot{300}{160}
{\ins{60pt}{90pt}{$v_0$}\ins{120pt}{100pt}{$v$}
\ins{100pt}{90pt}{$v'$}
\ins{120pt}{-5pt}{$h_v$}
\ins{235pt}{-5pt}{$N$}
\ins{255pt}{-5pt}{$N+1$}
}
{treelut2}{\label{n11} A Gallavotti tree 
}{0}
Let us consider the family of all trees which can be constructed
by joining a point $r$, the {\it root}, with an ordered set of $n\ge 1$
points, the {\it endpoints} of the {\it unlabeled tree}, 
so that $r$ is not a branching point. $n$ will be called the
{\it order} of the unlabeled tree and the branching points will be called
the {\it non trivial vertices}.
The unlabeled trees are partially ordered from the root to the endpoints in
the natural way; we shall use the symbol $<$ to denote the partial order. 
The number of unlabeled trees is $\le 4^n$, see eg  \S 2.1 of \cite{M5}.
The set of labeled  (or Gallavotti) trees $\TT_{h,n}$
are defined adding the above labels
\begin{enumerate}
\item We
associate a label $h\le N-1$ with the root and  
we introduce
a family of vertical lines, labeled by an an integer taking values
in $[h,N+1]$ intersecting all the non-trivial vertices, the endpoints and other points called trivial vertices.
The set of the {\it
vertices} $v$ of $\t$ will be the union of the endpoints, the trivial vertices
and the non trivial vertices. The scale label is $h_v$ and, if $v_1$ and $v_2$ are two vertices and $v_1<v_2$, then
$h_{v_1}<h_{v_2}$. $s_v$ is the number of subtrees with root $v$.
Moreover, there is only one vertex immediately following
the root, which will be denoted $v_0$ and can not be an endpoint;
its scale is $h+1$.  
\item To the end-points $v$ of scale $h_v\le N$ is associated $\tilde\LL V^{(h_v)}$; there is the constraint that the vertex $v'$ immediately preceding $v$, that is $h_{v'}=h_v-1$ is non trivial (as $\RR \LL=0$).
The end-points with $h_v\le N$  can be of type $\n$ or $Z$.
\item  To the end-points $v$ of scale $h_v= N+1$ is associated one of the terms  in $V^{(N+1)}$
\item Among the end-points, one distinguish between the normal ones, associated to terms not containing $J_\m, J_\m^5$, whose number is $\bar n=n-m$, 
and the others which are called special.
\item There is an $\RR$ operation associated to each vertex except the end-points and $v_0$; if the tree contributes to $\RR V^h$ it is associated
$\RR$ while if it contributes to $\b_h$ is associated $\LL$ and $s_{v_0}\ge 2$. 
\item A subtree with root at scale $k$ is called trivial if contains only the root 
and an endpoint of scale $k+ 1$
\end{enumerate}
%

The effective potential can be written as
\be
V^{(h)}(\sqrt{Z_h}\psi^{(\le h)},J,J^5) =
\sum_{n=1}^\io\sum_{\t\in\TT_{h,n}}
V^{(h)}(\t,\sqrt{Z_h}\psi^{(\le h)},J,J^5)\;,\ee
where, if $v_0$ is the first vertex of $\t$ 
and $\t_1,..,\t_{s_{v_0}}$
are the subtrees of $\t$ with root $v_0$,
$V^{(h)}$
is defined inductively by the relation, $h\le N-1$
\bea
&&V^{(h-1)}(\t,\sqrt{Z_h}\psi^{(\le h)},J,J^5)=\label{fr} \\
&&{(-1)^{{s_{v_0}}+1}\over {s_{v_0}}!} \EE^T_{h}[\bar
V^{(h)}(\t_1,\sqrt{Z_h}\psi^{(\le h)},J,J^5);..; \bar
V^{(h)}(\t_{{s_{v_0}}},\sqrt{Z_h}\psi^{(\le h)},J,J^5)]\nn\eea
where $\EE^T_{h}$ is the truncated expectation
with propagator $g_{i}^{(h)}$
and
\begin{itemize}
\item if $\t_i$ is non trivial
$\bar
V^{(h)}(\t_i,\sqrt{Z_h}\psi^{(\le h)},J,J^5)=
\RR V^{(h)}(\t_i,\sqrt{Z_h}\psi^{(\le h)},J,J^5)$
\item if $\t$ is trivial it is equal to one of the terms
in $\tilde \LL V^{(h)}$ if $h<N$, or to the one of the terms in $V^{(N+1)}$
if $h=N$.
\end{itemize}
We can write therefore the kernels in \pref{ap222} as 
\be
W^{(h)}_{l,m}(\underline x,\underline z)=\sum_{n=1}^\io \sum_{\t\in \TT_{h,n}} W^{(h)}_{l,m}(\t,\underline x,\underline z)
\ee
It is also  convenient to write
\be
\TT_{h,n}=\TT^1_{h,n}\cup\TT^2_{h,n}
\ee
with $\TT^1_{h,n}$ is the subset of $\TT_{h,n}$
containing all the trees with only end-points associated to $\tilde\LL V^k$, while 
$\TT^2_{h,n}$ contains the trees with at least one end-point associated to $V^{N+1}$.
We define
\be
W^{i (h)}_{l,m}(\underline x,\underline z,\underline \g)
=\sum_{n=1}^\io\sum_{\t\in\TT^i_{h,n}} W^{(h)}_{l,m}(\t;\underline x,\underline z,\underline \g)
\ee
with $i=1,2$ and $W^{ (h)}_{l,m}=W^{1 (h)}_{l,m}+
W^{2 (h)}_{l,m}$. A similar decomposition can be done for 
\be
\b^h_\n=\sum_{n=1}^\io\sum_{\t\in\TT^2_{h,n}} \b^h_\n(\t)\quad\quad \b^h_\ZZ=\sum_{n=1}^\io\sum_{\t\in\TT^2_{h,n}}\b^h_\ZZ(\t)
\ee
In this case by the compact support of the propagator only trees
contributing to $\TT^2_{h,n}$ are present; the contribution from $\TT^1$ are "chain graphs" and the localization correspond in momentum space to setting $k=0$, and $\hat g^h(0)=0$. Finally we can write
\be
\Pi^5_{\m,\n,\r}=\sum_{h=-\io}^N \Pi^{5,1}_{h,\m,\n,\r}+\sum_{h=-\io}^N \Pi^{5,2}_{h,\m,\n,\r}\label{supsup}
\ee
with $\Pi^{5,i}_{h,\m,\n,\r}=
\sum_{n=1}^\io\sum_{\t\in\TT^i_{h,n}}  W^h_{0,3}(\t)$.
The following lemma holds, see App. 2.
\begin{lemma} There exists a constant $\e$ such that,
for $|\ZZ_{k}|\le e^{\e  ({a  M})^{2} }$,
$\max (|\n_N|,..,|\n_h|,(\l/ Ma)^2\le \e  $  than if $m\le 3$, $d$ is the distance
between any two coordinate
\be  
||d^s W^{j(h)}_{l,m}||
\le C^{l+m}  \g^{(4-(3/2)l-m-s) h}\g^{\th_j(h-N)} 
\e^{\max(l/2-1,1)} 
\label{bb}
\ee
with $\th_1=0$ and $\th_2=\th$ for a constant $\th= 1/2$; moreover 
\be
|\b^h_\n|\le \e \g^{\th (h-N)}\quad |\b^h_\ZZ|\le \e \g^{\th (h-N)}\label{sapo}
\ee
\end{lemma}
The bound is proven showing the convergence of the expansion in 
$\n_k,\l$ under a smallness condition which is independent from $h$.
Note that if we perform a multiscale integration setting $\LL=0$
then the condition would be that $\l\le \e_h$ with $\e_h$ going to zero a $h\to-\io$. The bound is similar to the one in Lemma 1.2, with 
the same "dimensional" factor  $\g^{(4-(3/2)l-m) h}$.
 
A crucial point is that the contributions from trees $\TT^2$, that is the terms obtained by the contraction of the irrelevant terms,
have a gain
$\g^{\th (h-N)}$ with respect to the dimensional bounds. This fact, and the bound 
\pref{sapo} with
$\ZZ_{h-1}=1+\sum_{k=h}^N \b^k_\ZZ$ implies
\be
|\ZZ_{-\io}-1|\le C 
 \e
\quad\quad
|\ZZ_{-\io}-\ZZ_h|\le C \e \g^{\th (h-N)} 
\label{sep12} 
\ee
that is the wave function and the vertex renormalization is bounded uniformly in $h$.
In addition we can rewrite \pref{tok}  as
\be
\n_{h-1,i}=\g^{-h}(\n_{N,i}+\sum_{k=h}^N \g^k \b^{(h)}_{\n,i})\label{tok}
\ee
We consider the system
\be
\n_{h-1,i}=\g^{-h}(-\sum_{k\le h} \g^k \b^{(h)}_{\n,i})\label{mklh}
\ee
We can regard the right side of \pref{mklh}
as a function of the whole sequence $\n_{k,i}$, which we can denote by
$\underline \n=\{\n_k\}_{k\le N}$
so that \pref{mklh}
can be read as a fixed point equation $\underline \n=T(\underline \n)$
on the Banach space of sequences $\n$ 
such that $||\n||=\sup_{k\le N} \g^{\th (k-N)} |\n_k|\le C \l^2 (M a)^{-2}
$. By a standard proof, see e.g.  App A5 of \cite{GM1}, it is possible to prove that
there is a choice of $\n_i$ such that the sequence is bounded for any $h$. With this choice
\be
|\n_h|\le C   \g^{\th (h-N)} \e\label{bb1}
\ee
This means that the $\n_{h,i}$ is bounded so that the condition required in Lemma 2.3 are fulfilled;
moreover is an easy consequence of the proof of Lemma 2.3 and of \pref{bb1} that the limit 
$L\to\io$ can be taken; the proof is standard, see App. E of  \cite{GM1}.
Finally
we can choose $Z^5_{i,s}=1+O(\e)$ so that
\be
Z^5_{i,s,-\io}=Z^{J}_{i,s,-\io}\label{M4}
\ee
We finally to apply the above results and get bounds for the three current function.
By \pref{supsup} and the bound 
\pref{bb} with $l=0, m=3,s=0$ we get
\be
|\sum_{h=-\io}^N \Pi^{5}_{h,\m,\n,\r}(x,y,0)|_{1}\le C \sum_{h=-\io}^N \g^{h}<C_N
\ee
hence the Fourier transform $\hat \Pi^{5}_{h,\m,\n,\r}(p_1,p_2)$ is continuous; in addition \pref{bb} with $l=0, m=3,s=1+\th/2$ and $j=2$
\be
|\sum_{h=-\io}^N (|x|^{1+\th/2}+|y|^{1+\th/2})  \Pi^{5,2}_{h,\m,\n,\r}(x,y,0)|_{1}\le C \sum_{h=-\io}^N \g^{-\th/2 h}\g^{\th(h-N)}
<\bar C_N
\ee
hence $\sum_{h=-\io}^N\hat \Pi^{5,2}_{h,\m,\n,\r}$ has continuous derivative.

Note that $\sum_{h=-\io}^N \hat \Pi^{5,1}_{h,\m,\n,\r}$ has a part from trees containing $\n_h$ end-points 
verifying \pref{bb1}, which by the above argument is again differentiable. We remain then with the contribution from trees with three end.points associated to $Z^5,Z^J,Z^J$.
We can write the propagator as
\be
g^{(h)}_{i,s,s'}(x,y)=
\d_ {s,s'}{1\over L^4}\sum_k {f_h(|k|_T)\over -i \s^s_\m k_\m} e^{i k(x-y)}+r^h_{i,s,s'}(x,y)\label{lap2}
\ee
where $r^h(x,y)$ is defined by the above equation as the difference; one can verify, again by integration by parts, that fpr any $K$
\bea
&&|g^{(h)}_{i,s,s'}(x,y)|\le  \d_ {s,s'}{1\over Z_{h,i,s}}\g^{3(h+1)} {C_K \over 1+( \g^{h+1}  |x-y|_{\tilde T})^K}\nn\\
&& |r^{(h)}_{i,s,s'}(x,y)|\le \g^{3(h+1)}\g^{h-N}
 {C_K \over 1+( \g^{h+1}  |x-y|_{\tilde T})^K}
\label{lap1}
\eea
The above decomposition says that the lattice propagator is equal to the continuum one up to a term with a similar decay with an extra 
$\g^{h-N}$. Again the contribution of such terms is differentiable and finally we can replace the $\ZZ_h$ terms in  
 $\sum_{h=-\io}^N \hat \Pi^{5,1}_{h,\m,\n,\r}$ with $\ZZ_{-\io}$ up again to differentiable terms, by 
\pref{sep12}. In conclusion we get, see Fig. 5
\be
\hat \Pi_{\m,\r,\s}(p_1,p_2)=\hat \Pi^a_{\m,\r,\s}(p_1,p_2)
+\hat R_{\m,\r,\s}(p_1,p_2)\label{fon}
\ee
with, $p=p_1+p_2$,
 
\bea
&&\hat \Pi^a_{\m,\r,\s}(p_1,p_2)=\sum_{h_1\atop h_2,h_3} \sum_{i,s} \tilde \e_i \e_s  Q_{i}^3 { Z^{5}_{-\io,i,s}
\over Z_{-\io,i,s}} {Z^{J}_{-\io, i,,s}\over Z_{-\io,i,s}}
{Z^{J}_{-\io, i, s}\over Z_{-\io,i,s}}\nn\\
&&\int {dk \over (2\pi)^4}{\rm Tr}{f_ {h_1}(k)\over i  \s^{s}_\m k_\m}
i\s^{s}_\m{ f_{h_2}\over i \s^{s}_\m (k_\m+p_\m)} i \s^{s}_\n {f_ {h_3}\over i   \s^{s}_\m (k_\m+p^2_\m)}(i \s^{s}_\r)
\eea
\pref{fon} says that the Fourier transform of the 3-current correlation can be decomposed 
in the sum of two terms; the first $\hat \Pi^a_{\m,\r,\s}(p_1,p_2)$
is continuous and is a sum of triangle graphs
equal to the its analogue in the non-interacting continuous case
with momentum regularization, with
vertex and wave function renormalizations depending on the species and chirality.
The second $\hat R_{\m,\r,\s}(p_1,p_2)$ is a complicate series of terms which is differentiable.

The renormalizations in $\hat \Pi^a_{\m,\r,\s}(p_1,p_2)$
are however the same appearing in the 2-point and vertex correlations so that we can use the Ward Identities;
we can write, see App.2
\be
\hat S_{i,s}(k)={1\over (i \s^s_\m k_\m)}
({I \over Z_{i, s,-\io}}+r_1(k))\label{all1}\ee
and
\be
\hat \G_{\m,i,s}(k,p)={1\over (i \s^s_\m k_\m)}
{Z^J_{i,s,-\io}
\over Z^2_{i,s,-\io}} (i \s_\m^s +r_{2,\m}(k,p))
{1\over (i \s^s_\m (k_\m+p_\m))}\label{all}
\ee
with $|r_1(k)|\le C (a |k|)^\th$ and $|r_{2,\m}(k,p)|\le C (a |k|)^\th$
with $|p|\le |k|$.
\insertplot{460}{89}
{\ins{50pt}{40pt}{$=$}
\ins{130pt}{40pt}{$+$}
}
{figjsp44a}
{\label{h2} Graphical representation of \pref{fon} 
} {0}
By inserting \pref{all1}, \pref{all} in the Ward Identities \pref{wia} we get exact relations between the wave and vertex renormalizations, that is
\be {Z_{-\io,i,s}^{J}
\over Z_{-\io,i,s}}=1\ee
Note the crucial fact that the contribution from the terms $r_{i,\m}$, coming from the trees $\TT^2$,
is subleading.
In conclusion, we get 
\be
\hat\Pi^5_{\m,\r,\s}=\hat I_{\m,\r,\s}+\hat \RR_{\m,\r,\s}
\ee
with $\hat \RR$ with Holder continuous derivative and 
\be
\hat I_{\m,\r,\s}(p_1,p_2)=
(\sum_i \tilde \e_i Q_i^3)
\int {dk \over (2\pi)^4}{\rm Tr} {\chi(k)\over \not k }\g_\m \g_5 {\chi(k+p)\over \not k+
\not p}\g_\n{\chi(k+p^2)\over \not k+\not p^2}\g_\s
\ee
Note that $\hat I_{\m,\r,\s}(p_1,p_2)$ is the anomaly for non-interacting relativistic continuum fermions 
with a momentum regularization which
violates the vector current conservation, see \cite{BGM1}, \S 3.6 for the explicit computation
\be 
\sum_\m (p_{1,\mu} + p_{2,\mu}) \hat I_{\mu,\nu,\sigma}
 = {(\sum_i \tilde \e_i Q_i^3)\over 6\pi^{2}} p_{1,\alpha} p_{2,\beta} \varepsilon_{\alpha\beta\nu\sigma}\quad
\sum_\n p_{1,\nu} \hat I_{\mu,\nu,\sigma}={(\sum_i \tilde \e_i Q_i^3)\over 6\pi^{2}} p_{1,\alpha} p_{2,\beta} \varepsilon_{\alpha\beta\mu\sigma}
\ee
 up to 
$O(a^\th |\bar p|^{2+\th})$ corrections. 
In contrast with $\hat I_{\m,\r,\s}$, we have that $\hat\RR_{\m,\r,\s}$ has not a simple explicit expression, being expressed in terms of a convergent series depending on all the lattice and interaction details. However we use the differentiability of 
$\hat\RR_{\m,\r,\s}(p_1,p_2)$ to expand it at first order
obtaining, again up to  
$O(a^\th |\bar p|^{2+\th})$ corrections, using the Ward Identity 
%
%
\be
\frac{1}{6\pi^{2}}
(\sum_i \tilde \e_i Q_i^3)  p_{1,\alpha} p_{2,\beta} \varepsilon_{\alpha\beta\mu\sigma} +\sum_\n p_{1,\nu} \Big(\hat\RR_{\mu,\nu,\sigma}({0}, 
{ 0})\nn\\
 +\sum_{a=1,2}
\sum_{\rho} p_{a,\rho} 
\frac{\partial \hat\RR_{\mu,\nu,\sigma}}{\partial p_{a,\rho}}({0}, {0}) \Big)=0
\ee
This implies that \be \hat\RR_{\mu,\nu,\sigma}({0}, 
{ 0})=0\ee and
\be
{\partial \hat\RR_{\mu,\nu,\sigma}\over \partial p_{2,\beta}} = - {1\over 6\pi^{2}}\varepsilon_{\nu\beta\mu\sigma} (\sum_i \tilde \e_i Q_i^3)\quad\quad
{\partial\hat\RR_{\mu,\nu,\sigma}\over \partial p_{1,\b}} (0, {0}) = {1\over 6\pi^{2}}\varepsilon_{\n\beta\mu\s} (\sum_i \tilde \e_i Q_i^3)
\ee
Finally using such values we get
\bea
&&\sum_\m (p_{1,\mu} + p_{2,\mu})\hat\Pi^5_{\mu,\n,\s}(p_1,p_2)=\sum_{\a,\b}\frac{(\sum_i \tilde \e_i Q_i^3)}{6\pi^{2}} p_{1,\alpha} p_{2,\beta} \varepsilon_{\alpha\beta\nu\sigma}\\
&&+\sum_{\mu,\b}  
(p_{1,\mu} + p_{2,\mu})
({\hat \RR_{\mu,\nu,\sigma}\over \partial p_{2,\beta}}(0,0) p_{2,\beta}
+{\hat \RR_{\mu,\nu,\sigma}\over \partial p_{1,\beta}} (0,0)
p_{1,\beta})\nn
\eea
and the second term in the r.h.s. is 
\be
-\frac1{6\pi^2} (p_{1,\mu} + p_{2,\mu}) \sum_{a=1,2}(-1)^a  p_{a,\beta}\varepsilon_{\nu\beta\mu\sigma}(\sum_i \tilde \e_i Q_i^3)=
\frac1{3\pi^2} 
p_{1,\mu} p_{2,\beta}\varepsilon_{\nu\beta\mu\sigma}(\sum_i \tilde \e_i Q_i^3)
\ee
which implies the Theorem 1.1 \qed

\section{Appendix 1: truncated expectations}

\subsection{The Brydges-Battle-Federbush formula} 

The starting point is the formula
\be
\EE_A(\prod_ {i=1}^n e^ {i \e_i \a_i A_{\m_i}(x_i)})= e^ {-{1\over 2}\sum_ {i,j}  \e_i \e_j \a_i \a_j g^A_{\m_i,\m_j}(x_i,x_j) }\label{sim}
\ee
Let us define
\be
e^{-V}\equiv e^{-{1\over 2} \sum_{j,j'\in X} \bar V_{j,j'}    }
\ee
with
$X=(1,2,..,n)$ and $
\sum_{i,j\in X} \bar V_{i,j}=\sum_{i\le j}V_{i,j}$
$\bar V_{i,i}=V_{i,i}$ and $V_{i,j}=(\bar V_{i,j}+\bar V_{j,i})/2$. 

The connected part $e^{-V(X)}|_T$ (corresponding to the truncated expectation) verify
\be
e^{-V(X)}=\sum_\pi \prod_{Y\in \pi}e^{-V(Y)}|_T\label{inc}
\ee
where the sum is over $\pi$ are the partitions of $X$, that is $Y_1,Y_2,...$
with $Y_1\cup Y_2\cup..=X$. 

If $X_1=\{1\}$ we can define
\be
W_X(X_1;t_1)=\sum_{\ell} t_1(l) V_l
\ee
where $\ell=(j,j')$ is a pair of elements $j,j'\in X$
and $ t_1(l)=t_1$ if $l$ crosses the boundary of $X_1$ ($\partial X_1$),
that is if it connect $1$ with $j\not= 1$;  $t_1(\ell)=1$ otherwise. More explicitely
\bea
&&W_X(X_1,t_1)=V_{1,1}+t_1 \sum_{k\ge 2} V_{1,k}
+\sum_{2\le k\le k'} V_{k,k'}=\nn\\
&&t_1 (V_{1,1}+\sum_{k\ge 2} V_{1,k}  +\sum_{2\le k\le k'} V_{k,k'})
+(1-t_1)(V_{1,1}+\sum_{2\le k\le k'} V_{k,k'})   =\nn\\
&&
t_1 V(X)+(1-t_1) (V(X_1)+V(X/X_1))\eea
We get $W_X(X_1,0)=V(X_1)+V(X/X_1)$, that is if $t_1=0$ $X_1$ is disconnected from the rest. Therefore, using that
$
\partial_1 W(X_1,t_1)=\sum_{k\ge 2} V_{1,k}=\sum_{l_1} V_{l_1}$ 
we can write
\be
e^{-V(X)}=\int_0^1 dt_1 \partial_1 e^{-W_X(X_1,t_1)}+e^{-W_X(X_1,0)}
\ee
and
\be
e^{-V(X)}
=\int_0^1 
dt_1 \sum_{k\ge 2} V_{1,k} e^{-W_X(X_1,t_1)}+ e^{-V(X_1)} e^{-V(X/X_1)}\label{tri}
\ee
We have therefore expressed 
$e^{-V(X)}$ as the sum of two terms; in the first there is a bond  $(1,k)$ between
$X_1$ and the rest is found, in the second $X_1$ is decoupled.
If $n=2$ the first term is the connected part. 

If $n\not=2$
we further decompose the first term in the r.h.s of \pref{tri};
we write $X_2=\{1,k\}$ and
\bea
&&\int_0^1 dt_1 \sum_{k\ge 2} V_{1,k} 
e^{-W_X(X_1,t_1)}=\\
&&\int_0^1 dt_1 \sum_{k\ge 2} V_{1,k}
\int_0^1 dt_2 \partial_{t_2} e^{-W_X(X_1,X_2;t_1,t_2)}+\int_0^1 dt_1 \sum_{k\ge 2} V_{1,k}
e^{-W_X(X_1,X_2;t_1,0)}\nn
\eea
where
\be
W_X(X_1,X_2,t_1,t_2)=(1-t_2)[ W_{X_2}(X_1,t_1)+V(X/X_2)] +t_2 W_X(X_1,t_1)
\ee
and for $X_2=(1,2)$
\be
W_X(X_1,X_2,t_1,t_2)=
V_{1,1}+V_{2,2}+  t_1 t_2 \sum_{k\ge 3} V_{1,k}+
t_1 V_{1,2}
+t_2
 \sum_{k\ge 3} V_{2,k}
+\sum_{3\le k\le k'}  V_{k,k'}\label{botr}
\ee
Suppose that $X=\{1,2,3\}$ and $X_2=\{1,2\}$, then
$
W_{X_3}(X_1,X_2,t_1,t_2)=
V_{1,1}+V_{2,2}+  t_1 t_2 V_{1,3}+
t_1 V_{1,2}
+t_2 V_{2,3}+V_{3,3}$ and
\bea
&&\int_0^1 dt_1 V_{1,2} 
e^{-W_X(X_1,t_1)}=\label{ill}\\
&&\int_0^1 dt_1 V_{1,2}
\int_0^1 dt_2 (t_1 V_{1,3}+V_{2,3})
 e^{-W_X(X_1,X_2;t_1,t_2)}+[\int_0^1 dt_1 V_{1,2}
e^{-W_{X_2}(X_1;t_1)}]e^{-V(X/X_2)}\nn
\eea
and the first term is connected; similar expressions for $X_2=\{1,3\}$.

Proceeding in this way
\bea
&&e^{-V(X)}=\sum_{r=1}^n \sum_{X_r\subset X}\sum_{X_1,..,X_{r-1}}
\sum_T[\prod_{\ell\in T} V_l]\nn\\ 
&&[\sum_{X_1,..,X_{r-1} }\int_0^1 dt_1...\int_0^1 dt_{r-1}
\prod_{\ell\in T}{\prod_{k=1}^{r-1} t_k(\ell)\over t_{n(\ell)}} e^{-W_{X_r}(X_1,..,X_{r-1};t_1,..,t_{r-1})}] 
e^{-V(X/X_r)}
\label{issaa}
\eea
where $X_1\subset X_2\subset...X_{r-1}$
are sets such that $|X_i|=i$,
$T$ is a tree composed by $r-1$ lines $\ell=(j,j')$ such that all the
boundaries $\partial X_k$ are intersected at least by a line $\ell=(j,j')$,
\be
W_X(X_1,..,X_r;t_1,..,t_r)=\sum_{l} t_1(l)t_2(l)...t_r(l) V_l\label{ben11}
\ee
with $t_i(l)=t_i$ if $l$ crosses $\partial X_i$ and $t_i(l)=1$ otherwise, $n(l)$ is the max over $k$ such that $l$ crosses $\partial X_k$.
For instance in the case \pref{ill}
the trees are $l_1=(1,2), l_2=(2,3)$ so that 
$t_1(l_1)=t_1$, $t_1(l_2)=1$, $t_2(l_2)=t_2$; 
and $l_1=(1,2)$, $l_2=(1,3)$ so that $t_1(l_1)=t_1$ and 
$t_1(l_2)=t_1$,$ t_2(l_2)=t_2$.

We can reverse the sum over $T$ and $X$
\be
\sum_T \sum_{X_1,..,X_{r-1}}=\sum_{X_1,..,X_{r-1}}\sum_T
\ee
where in the l.h.s. the sets have to be compatible with $T$.
If $n'(\ell)$ is the minimal $k$ such that $\ell$ crosses $X_k$
we have ${\prod_{k=1}^{r-1} t_k(\ell)\over t_{n(\ell)}}=
t_{n'(\ell)}...t_{n(\ell)-1}$ and, see e.g. Lemma 2.3 in \cite{M5}
\be
\sum_{X_1,..,X_{r-1}\atop fixed T}\int_0^1 dt_1...\int_0^1 dt_{r-1}
t_{n'(\ell)}...t_{n(\ell)-1}=1
\ee
By calling
\be
dp_T(t)=\sum_{X_1,..,X_{r-1}\atop fixed T}
{\prod_{k=1}^{r-1} t_k(l)\over t_{n(l)}}
\ee 
we get
\be
e^{-V(X)}|_T=\sum_T [\prod_{\ell\in T} V_l] \int_0^1 d\underline t
dp_T(t)
\e^{-\sum_{\ell\in X'\atop \ell \not T} t_{n'(\ell)}...t_{n(\ell)-1 }V_{\ell}}
\ee
where $\ell\in X$ means $j,j'\in (1,..,n)$.

\subsection{The Gawedzki-Kupiainen-Lesniewski formula}

We can write the simple expectations as
\be
\EE(\tilde\psi(P_1)...\tilde\psi(P_r))=\int \prod_{i,j}  d\h_{i,j} e^{-\sum_{j,j'} V_{jj'}  }
\ee
with $V_{jj'}=\sum_{i=1}^{|P_j|}\sum_{i'=1}^{|P_j'|} \h^+_{x_{ij} } g(x_{ij}, x_{i'j'})  \h^-_{x_{i'j'}}$
and $\h^\pm_{i,j}$ is a set of Grassmann variables. Again we can write 
$e^{-\sum_{j,j'} V_{jj'}}$ as in \pref{issaa} obtaining
\bea
&&\EE^T(\tilde\psi(P_1)...\tilde\psi(P_r))=\\
&&\int \prod d\h^+_{i,j}d\h^-_{i,j}
 \sum_T[\prod_{l\in T} V_l]\int_0^1 \underline d\bar t
dp_T(\underline t)
\e^{-\sum_{\ell\in X'} t_{n'(\ell)}...t_{n(\ell)-1 }V_{\ell}}
\eea
with  $V_{\ell}=\sum_{i}\sum_{i'} \h^+_{i,j} g(x_{ij},x_{i'j' }  \h^-_{i,j}$, $\ell=(j,j')$.
For each tree $T$
we divide the $\h$ in the ones appearing in $T$, called $\tilde \h$, and the rest, called $\bar \h$ so that, if $\sum_{\ell\in X'} t_{n'(\ell)}...t_{n(\ell)-1 }V_{\ell}=\tilde V(\underline t)+\bar V(\underline t)$ with 
$\bar V(\underline t)$ obtained setting  $\tilde \h=0$
\be
\EE^T(\tilde\psi(P_1)...\tilde\psi(P_r))=\sum_T[\prod_{l\in T} g_\ell ]\int_0^1 \underline d\bar t
dp_T(\underline t)\int \prod d\bar\h^+_{i,j}d\bar\h^-_{i,j}
\e^{-\bar V(\underline t)}
\ee
and $\prod d\bar\h^+_{i,j}d\bar\h^-_{i,j}
\e^{-\bar V(\underline t)}=\det G_T$ with $G_T$ with elements 
$t_{n'(j,j')}...t_{n(jj')-1 }g(x_{ij},x_{i'j' })$. Fixed $T$ we can relabel the $X_k$ so that $t_{j}...t_{j'-1 }=u_j u_{j'}$
with $u_1=v_1$, $u_j=t_{j-1} u_{j-1} +v_j \sqrt{1-t^2_{j-1}}$ with $v_j$ 
orthonormal, and $u_1 u_2=t_1$, $u_1 u_3=t_1 t_2$, $u_2 u_3=t_2$ and so on.

\section{Appendix 2: proof of lemma 2.3}

The proof is a generalization of the proof of lemma 2.2 adapted to the tree structure.
We define 
$P_v$ as the set of field labels of the external fields of $v$ 
and if
$v_1,\ldots,v_{s_v}$ are the $s_v$
vertices immediately following $v$, we denote by $Q_{v_i}$ the intersection of $P_v$ and
$P_{v_i}$. This definition implies that $P_v=\cup_i Q_{v_i}$. The
union of the subsets $P_{v_i}\bs Q_{v_i}$ are the internal fields of $v$.
The set of all $P_v$, $v\in \t$ is called $\PP$, and the set of all $P_v$ with $v\ge \t_i$ is called $\PP_i$.
From \pref{fr} we get, if $n_{v_0}$ is the number of coordinate
\be
V^{(h)}(\t)=\sum_{\PP}
a^{4 n_{v_0}} \sum_{x_{v_0}}  W^{(h)}_{\t,\PP}(x_{v_0})
[\prod_{f\in P_{v_0}}\sqrt{Z_h}
\psi^{\e(f)(\le h)}_{x(f),i(f),s(f)}][\prod_f J(x_f)]
\ee
By definition we have a truncated expectation associated to each $v$ in the tree $\t$ non associated to an end-point; we 
can write each of them by 
the Gawedzki-Kupiainen-Lesniewski formula. The $\RR$ operation is applied
and by an iterative procedure and the number of zeros 
associated to propagators of $T$ and
and the derivative on the fields are bounded by a constant; see e.g. \S 3 
of \cite{BM}. 

The bound is done using 
the Gram bound for the determinant;
to each vertex is therefore associated a spanning tree $T_v$ which is used to perform the sum over the coordinate difference, and $T=\cup_v T_v$.
The sum over coordinates of the propagators in $T$ and the estimates of the determinants give a factor
$\g^{-4 h_v (s_v-1)}
\g^{3/2 h_v (\sum_i |P_{v_i}|-|P_v|)}$, if $S_v$ is the number of subtrees 
with root $v$. 
The renormalization produces a factor $ \prod_v \g^{-z_v(h_v-h_{v'}  ) }$ is produced by the $\RR$ operation and
$z_v=2$ if $|P_v|=2$ and there are no $J$ fields, $z_v=1$ if $|P_v|=2$ and there is a single $J$ field, $z_v=0$ otherwise. 
To the end-points with $i$ $\psi$ fields and $j$ $J$ fields is associated by lemma 2.1
a factor $ \g^{(4-3 i_v/2-j_v)N} (\l^{i_v/2} (a  M)^{2-i_v})$
with $(4-3 i_v/2-j_v)<0$ and $i_v\ge 4$ and $(a  M)^{2-i_v}<(a  M)^{-2}$. We get therefore
\bea
&&a^{4 n_{v_0}} \sum_{x_{v_0}} |W_{\t,\bP}(x_{v_0})|\le L^4 \sum_T \prod_{v\, not\, e.p.} {1\over s_v!} 
C^{\sum_{i=1}^{s_v}|P_{v_i}|-|P_v|} \g^{-4 h_v (s_v-1)}\\
&&
\g^{3/2 h_v (\sum_i |P_{v_i}|-|P_v|)}[\prod_v \g^{-z_v(h_v-h_{v'}  ) } ]
[\prod_{v\;e.p. \; not\; \n,Z}
 \g^{(4-3 i_v/2-j_v)N} ][\prod_{v  \;e.p.\;\n} \g^{h_v} ] 
\e^{\bar n}\nn
\eea
By using that 
\bea
&&\sum_v (h_v-h)(s_v-1)=\sum_v (h_v-h_{v'})(\sum_{i,j } m^{i,j}_v-1)\nn\\
&&\sum_v (h_v-h) (\sum_i |P_{v_i}|-|P_v|)=\sum_v (h_v-h_{v'})(\sum_{i,j} i  m^{i,j}_v-|P_v|)
\eea
where $m^{i,j}_v$ is the number of end-points following $v$  with $i$ $\psi$ fields and $j$ $J$ fields , we get
\bea
&&a^{4 n_{v_0}} \sum_{x_{v_0}} |W_{\t,\bP,T}(x_{v_0})|\le L^4 
\g^{-h[-4+{3|P_{v_0}|\over 2}- \sum_{i,j} (3 i/2-4) m^{i,j}_{v_0}]} 
\e^{\bar n}
\nn\\
&&\prod_{v\; not \; e.p.} \left\{ {1\over s_v!}
C^{\sum_{i=1}^{s_v}|P_{v_i}|-|P_v|}
\g^{-(-4+{3|P_v|\over 2}-\sum_{i,j}(3 i/2-4) m^{i,j}_v+z_v)
(h_v-h_{v'})}\right\}\nn\\
&& [
\prod_{v\;e.p. \; not\; \n}\g^{(4-3 i_v/2-j_v)N}][\prod_{v\; e.p.\; \n} \g^{h_v}]
\nn\eea
We use now that
\be
\g^{h \sum_{i,j } m^{i,j}_{v_0}}\prod_{v\; not \; e.p.} \g^{\sum_{i,j }(h_v-h_{v'}) m^{i,j}_v}=\prod_{v \; e.p.} \g^{h_{v^*}}
\ee
where $v^*$ is the first non trivial vertex following $v$; this implies
%
%
\be
\g^{h\sum_{i,j} (3 i/2-4) m^{i,j}_{v_0}}\prod_{v\; not \; e.p.} \g^{\sum_{i,j}(3 i/2-4) m^{i,j}_v(h_v-h_{v'}) }
=\prod_{v\; e.p\; not\; \n} \g^{h_{v^*} (3 i_{v}/2-4)}\prod_{v \;  e.p.\; \n} \g^{-h_v}
\ee
so that
\bea
&&a^{4 n_{v_0}} \sum_{x_{v_0}}  |W_{\t,\bP,T}(x_{v_0})|\le L^4 
\g^{-h[-4+{3|P_{v_0}|\over 2}]} 
\prod_{v\; e.p\; not\; \n} \g^{h_{v^*} (3 i_{v}/2-4)} \e^{\bar n} \nn\\
&&\prod_{v\; not\; e.p.} \left\{ {1\over s_v!}
C^{\sum_{i=1}^{s_v}|P_{v_i}|-|P_v|}
\g^{-(-4+{3|P_v|\over 2}+z_v)
(h_v-h_{v'})}\right\} [
\prod_{v\;e.p. \; not\; \n}\g^{(4-3 i_v/2-j_v)N}]
\eea
Finally we use the relation
\be
[\prod_{v\; e.p.} \g^{h_{v^*} j_v}][ \prod_{v \; e.p. } \g^{-h_{v^*} j_v}]=[\prod_{v\; e.p.} \g^{h_{v^*} j_v}]
\g^{-h \sum_{i,j } j m^{i,j}_{v_0}}\prod_{v\; not \; e.p.} \g^{-\sum_{i,j }(h_v-h_{v'}) j m^{i,j}_v}
\ee
and using that $\sum_{i,j}   j m^{i,j}_v  =n^J_v$ we finally get  ($j_v=0$ if $v$ is a $\n$-e.p.) 
\bea
&&a^{4 n_{v_0}} \sum_{x_{v_0}}  |W_{\t,\bP,T}(x_{v_0})|\le L^4 
\g^{-h[-4+{3|P_{v_0}|\over 2}+n^J_{v_0}  ]} 
\e_h^{\bar n}
\nn\\
&&\prod_{v\,\hbox{\ottorm not e.p.}} \left\{ {1\over s_v!}
C^{\sum_{i=1}^{s_v}|P_{v_i}|-|P_v|}
\g^{-(-4+{3|P_v|\over 2}+z_v+n^J_v )
(h_v-h_{v'})}\right\} [
\prod_{v\;e.p. \; not\; \n,Z}\g^{(4-3 i_v/2-j_v)(N-h_{v^*})}]
\nn\eea
In conclusion
\be
a^{4 n_{v_0}} \sum_{x_{v_0}}  |W_{\t,\bP,T}(\xx_{v_0})|\le L^4 
\g^{-h d_{v_0}} C^n \e^{\bar n}
[\prod_{\tilde v} 
{1\over s_{\tilde v}!}\g^{-d_{\tilde v} (h_{\tilde v}-h_{\tilde v'}) }
][
\prod_{v\;e.p. \; not; \n,Z}\g^{(4-3 i_v/2-j_v)(N-h_{v^*})}]\label{ess}
\ee
where: $\tilde v\in \tilde V$ are the vertices on the tree such that $\sum_i |P_{v_i}|-|P_v|\not=0$, $\tilde v'$ is the
vertex in $\tilde V$ immediately preceding $\tilde v$ or the root;
$d_v=-4+{3|P_v|\over 2}+n^J_v+z_v$.
Finally the number of addenda in $\sum_{T\in {\bf T}}$ is bounded by
$\prod_{v} s_v!\;
C^{\sum_{i=1}^{s_v}|P_{v_i}|-|P_v|}$, see e.g.\S 2.1 of \cite{GM} . 
In order to bound the sums over the scale labels and $\bP$ we first use
the inequality
\be
\prod_{\tilde v} \g^{-d_{\tilde v} (h_{\tilde v}-h_{\tilde v'}) }
\le [\prod_{\tilde v} \g^{-{1\over 2}(h_{\tilde v}-h_{\tilde v'})}]
[\prod_{\tilde v}\g^{-{3|P_{\tilde v}|\over 4}}]
\ee
where $\tilde v$ are the non trivial vertices, and $\tilde v'$ is the
non trivial vertex immediately preceding $\tilde v$ or the root. The
factors $\g^{-{1\over 2}(h_{\tilde v}-h_{\tilde v'})}$ in the r.h.s.
allow to bound the sums over the scale labels by $C^n$ and $\sum_{\bf P} \prod_{\tilde v}\g^{-{3|P_{\tilde v}|\over 4}}\le C^n$, see \S 3.7 of
of \cite{M5} .

Let us consider the improvement of the bound.
If $\TT^*$ is the set of trees with at least an end-point not of $\n, Z$ type then, for $0<\th<1$
\be
\sum_{\t\in \TT^*}
\sum_{\bP,T}
a^{4 n_{v_0}} \sum_{x_{v_0}}  |W_{\t,\bP,T}(x_{v_0})|\le L^4 \g^{(4-(3/2)l-m) h} \g^{\th (h-N)}
\e^{\max(l/2-1,1)} \label{lus}
\ee
To prove \pref{lus}
let be $\hat v$ the non trivial vertex following an end-point
not of $\n, Z$ type; hence we can rewrite in \pref{ess} 
\be
[\prod_{\tilde v}
\g^{-d_{\tilde v} (h_{\tilde v}-h_{\tilde v'}) }] =
[\prod_{\tilde v}
\g^{-(d_{\tilde v}-\th) (h_{\tilde v}-h_{\tilde v'}) }] \g^{\th (h-h_{\hat v}) }
\ee
and 
\be
\g^{\th (h-h_{\hat v}) }[
\prod_{v\;e.p. \; not\; \n,Z}\g^{(4-3 i_v/2-j_v)(N-h_{v^*})}]\le  \g^{\th (h-N) }
\ee
as $\prod_{v\;e.p. \; not\; \n, Z}\g^{(4-3 i_v/2-j_v)(N-h_{v^*})}\le  \g^{-\th (N-h_{\hat v}) }$ as there is at least an end-point not $\n,Z$.
Noting that $d_{\tilde v}-\th>0$ one can perform the sum as above, and the same bound is obtained with an extra
$\g^{\th (h-N)}$. \qed

\vskip.3cm
In presence of a $\phi$ term there is a new relevant coupling proportional to $\psi\phi$, whose local part  is vanishing
again by the compact support of the propagator. We can compare the bound from the one of a term of the effective potential with $l=2$
with two $\n$ end-points.
On each tree there is a vertex which is the root of the subtree
to which belong both the end-points associated with $(\psi\phi)$; there is an integral missing giving an extra factor $\g^{2 \bar h}$
reproducing the similar factor associated to the $\n$ end-points. There is a decay factor proportional to $x-y$ at scale $\g^{\bar h}$
and, from the trees beloging to $\TT^*$, an extra $\g^{\th (h-N)}$; see e.g. \S 3.D of \cite{M4a}. A similar argument
holds for the vertex function. Finally the proof of the 
$L\to\io$ limit is an easy corollary of the proof of lemma 2.3, 
see e.g.  App D of \cite{M4a}.

\section{Appendix 3: symmetries}

By symmetry there are no quadratic contributions with $i'\not=i$.
There is invariance under the transformation $\psi^\pm_{k,s}\to \e_s\psi^\mp_{\tilde k,s}\s_1$, $J_{\kk}\,A_\kk\to J_{\tilde\kk},A_{\tilde\kk}$ invariant, 
if $\tilde k$ is equal to $k$ with $k_0,k_1$ replaced with $-k_0,-k_1$ 
and $k_2,k_3$ invariant.
As $j= (2,3)$ $\s_1 \s_j \s_1=-\s_j$  hence $\sum_k \sin k_j \psi^+ _{k,s}\s_j\psi^-_{k,s}\to \sum_k \sin k_j \psi^-_{\tilde k,s}\s_1 \s_j\s_1\psi^+_{\tilde k,s}=\sum_k \sin k_j \psi^+ _{k,s}\s_0\psi^-_k$
and for $j=0,1$  $\s_1 \s_j \s_1=\s_j$ hence
$\sum_k \sin k_j \psi^+ _{k,s}\s_0\psi^-_k\to \sum_k \sin k_j \psi^-_{\tilde k,s}\s_1 \s_j\s_1\psi^+_{\tilde k,s}=\sum_k \sin k_j \psi^+ _{k,s}\s_j\psi^-_{k,s}$;
and
$\sum_k \cos k_j \psi^+ _{k,L}\s_0\psi^-_{k,R}\to \sum_k -\cos k_i \psi^-_{\tilde k,L}\s_1 \s_0\s_1\psi^+_{\tilde k,R}$. Similarly 
there is invariance under the trasformation $\psi^\pm_{k,s}\to \e_s\psi^\mp_{\tilde k,s}\s_2$, $J_{\kk}\,A_\kk\to J_{\tilde\kk},A_{\tilde\kk}$ invariant, 
if $\tilde k$ is equal to $k$ with $k_0,k_2$ replaced with $-k_0,-k_2$ 
and $k_1,k_3$ invariant.
As $\s_2 \s_j \s_2=-\s_j$ $j= (1,3)$ hence $\sum_k \sin k_j \psi^+ _{k,s}\s_j\psi^-_{k,s}\to \sum_k \sin k_j \psi^-_{\tilde k,s}\s_1 \s_j\s_1\psi^+_{\tilde k,s}=\sum_k \sin k_j \psi^+ _{k,s}\s_0\psi^-_k$
and for $j=0,2$  $\s_2 \s_j \s_2=\s_j$ hence
$\sum_k \sin k_j \psi^+ _{k,s}\s_0\psi^-_k\to \sum_k \sin k_j \psi^-_{\tilde k,s}\s_1 \s_j\s_1\psi^+_{\tilde k,s}=\sum_k \sin k_j \psi^+ _{k,s}\s_j\psi^-_{k,s}$;
and
$\sum_k \cos k_j \psi^+ _{k,L}\s_0\psi^-_{k,R}\to -\sum_k \cos k_i \psi^-_{\tilde k,L}\s_1 \s_0\s_1\psi^+_{\tilde k,R}$. 

We can write $\sum_k k_2   \psi^+_{\tilde k,s}A_2\psi^-_{\tilde k,s}=\sum_k k_2 
[a \s_0+b_\m \s_1+c_\m\s_2+d_\m\s_3]$.
We apply the first transformation to
$\sum_k k_2 \psi^+_{k,s}\psi^-_{k,s}(a \s_0+b \s_1+c\s_2+d\s_3)\to -\sum_k k_2\psi^+_{\tilde k}\s_1 (a \s_0+b \s_1+c\s_2+d\s_3)
\s_1\psi^-_{\tilde k}= -\sum_k k_2\psi^+_{k,s}\psi^-_{k,s}
(a \s_0+b \s_1-c\s_2-d\s_3)
\s_1\psi^-_k
$ hence $a=b=0$. Now we apply the second transformation
then $\sum_k k_2\psi^+_{k,s}\psi^-_{k,s}
\s_2(c\s_2+d\s_3)\s_2\to  -\sum_k k_2\psi^+_{\tilde k,s}\psi^-_{\tilde k,s}
(c\s_2-d\s_3)=\sum_k k_2\psi^+_{\tilde k,s}\psi^-_{\tilde k,s}
(c\s_2-d\s_3)
$ hence $d=0$. Then $\sum_k k_2\psi^+_{\tilde k,s}A\psi^-_{\tilde k,s}
=\sum_k k_2 b  \psi^+_{k,s}\s_2\psi^-_{k,s}$, and the geeral relation follows from isotropy. Proceeding in a similar way with the terms with different chirality $\sum_k k_2 \tilde b  \psi^+_{\tilde k,L}\s_2\psi^-_{\tilde k,R}\to -\sum_k k_2 \tilde b  \psi^+_{\tilde k,L}\s_2\psi^-_{\tilde k,R}$ hence $\tilde b=0$. 

Finally by the first tranformation	
$ \sum_k \psi^+_{k,L}\psi^-_{k,R}(a \s_0+b \s_1+c\s_2+d\s_3)\to 
\sum_k \psi^+_{\tilde k,L}\s_1 (a \s_0+b \s_1+c \s_2+d\s_3)
\s_1\psi^-_{\tilde k,R}= \sum_k \psi^+_{k,L}(a \s_0+b \s_1-c\s_2-d\s_3)
\s_1\psi^-_{k,R}$
so that $c=d=0$; by the second 
$\sum_k \psi^+_{k,L}\psi^-_{k,R}(a \s_0+b \s_2)\to 
\sum_k \psi^+_{\tilde k,L}\s_2 (a \s_0+b \s_1)
\s_2\psi^-_{\tilde k,R}=\sum_k \psi^+_{\,L}\s_2 (a \s_0-b \s_1)
\s_2\psi^-_{k,R}$ hence $b=0$.

\vskip.3cm
{\it Acknowledgements. } This work has been done partly at the Institute for Advanced Study, Princeton. We got support also from 
 MUR, project MaQuMA, PRIN201719VMAST01, and INDAM-GNFM.

\end{document}